\documentclass[11pt]{article} 
\pdfoutput=1

\usepackage{amsmath} 
\usepackage{amsthm} 
\usepackage{amssymb}	
\usepackage{graphicx} 
\usepackage{multicol} 
\usepackage{feynmp}
\usepackage[squaren,Gray]{SIunits}
\usepackage{jheppub}
\usepackage{subfigure}
\usepackage{multirow}
\usepackage{makecell}
\usepackage{bm}

\graphicspath{ {img/} }

\renewcommand{\v}[1]{\ensuremath{\mathbf{#1}}} 

\newcommand{\uv}[1]{\ensuremath{\mathbf{\hat{#1}}}} 
\newcommand{\avg}[1]{\left< #1 \right>} 
\newcommand{\pd}[2]{\frac{\partial #1}{\partial #2}} 
 
\let\baraccent=\= 
\renewcommand{\=}[1]{\stackrel{#1}{=}} 

\theoremstyle{definition}
\newtheorem{dfn}{Definition}
\newtheorem*{rst}{Result}
\theoremstyle{remark}

\newcommand{\gev}[1]{$\unit{#1}{\giga\electronvolt}$}
\newcommand{\tev}[1]{$\unit{#1}{\tera\electronvolt}$}

\newcommand{\gevm}[1]{\unit{#1}{\giga\electronvolt}}
\newcommand{\tevm}[1]{\unit{#1}{\tera\electronvolt}}

\newcommand{\half}{$\frac{1}{2}$\ }

\newcommand{\sigv}[1]{\avg{\sigma v}_{#1}}

\title{Dark matter and localised fermions from spherical orbifolds?}
\abstract{We study a class of six-dimensional models based on positive curvature surfaces (spherical 2-orbifolds) as extra-spaces. Using the Newman-Penrose formalism, we discuss the particle spectrum in this 
class of models. The fermion spectrum problem, which has been addressed with flux compactifications in the past, can be avoided 
using localised fermions. In this framework, we find that there are four types of geometry compatible with the existence of a stable dark matter 
candidate and we study the simplest case in detail. 
Using the complementarity between collider resonance searches and relic density constraints, we show that this class of models is under tension, unless the model lies in a funnel region characterised by a resonant Higgs s-channel in the dark matter annihilation.}
\author[a]{Giacomo Cacciapaglia,}
\author[a,1]{Aldo Deandrea,\note{Also Institut Universitaire de France, 103 boulevard Saint-Michel, 75005 Paris, France.}}
\author[a,b]{Nicolas Deutschmann}
\affiliation[a]{Universit\'e de Lyon, France; Universit\'e Lyon 1,
CNRS/IN2P3, UMR5822 IPNL,\\ F-69622 Villeurbanne Cedex, France.}
\affiliation[b]{Centre for Cosmology, Particle Physics and Phenomenology (CP3), Universit\'e catholique de Louvain,
Chemin du Cyclotron 2, B-1348 Louvain-la-Neuve, Belgium.}
\emailAdd{g.cacciapaglia@ipnl.in2p3.fr}
\emailAdd{deandrea@ipnl.in2p3.fr}
\emailAdd{n.deutschmann@ipnl.in2p3.fr}
 \preprint{LYCEN 2016-01}
\begin{document}
\maketitle
\section{Introduction}
Since the first proposals at the dawn of particle physics by Nordstr\"om \cite{Nordstrom:1988fi}, Kaluza \cite{Kaluza:1921tu} and Klein \cite{Klein:1926tv}, extra dimensional spaces have proven a useful playground for the exploration of extensions of the Standard Model (SM) of Particle Physics. Compact spaces, after the Kaluza-Klein (KK) reduction of the extra dimensional fields, naturally contain many massive replicas of the SM states, appearing as towers of KK resonances. In this paper, we are interested in the possibility that one (or more) of such states may play the role of Dark Matter. This idea is more than a decade old \cite{Servant2003a}, and it was a natural consequence of models where all the SM fields propagate in the extra dimension, as in Universal Extra Dimension (UED) incarnations \cite{Antoniadis:1990ew,Antoniadis:1998ig,Appelquist2000}. The stability of one of the KK modes derives from the presence of a residual discrete symmetry, left over from the breaking of the extended Poincare invariance by the compactification of the extra spatial dimensions.
UED models have been thoroughly investigated in the flat (zero-curvature) case, while
geometries with non-zero curvature are much less scrutinised even though they are an interesting alternative~\footnote{Models in warped space, or Randall-Sumdrum \cite{Randall:1999vf}, are an exception, as the asymmetric curvature eliminates the possibility of residual discrete symmetries. For an attempt to build double-throat models with KK Dark Matter, see \cite{Agashe:2007jb}.}. In all these models, 
it is common for the lightest Kaluza-Klein particle to be made stable in order to provide a candidate for the Dark Matter particle. 
The stability of Dark Matter is often imposed by adding symmetries to the model (typically a new parity). The most common situation is that the interactions in the bulk of the extra space respect the parity, while the presence of fixed points where lower dimension interactions can be added violates it: one is therefore left with the only option of imposing the parity invariance by hand on the localised interactions. It was later realized in \cite{Cacciapaglia:2009pa} that the requirement of a compact space that preserves the Dark Matter parity, including the fixed points, is a powerful selection rule on the choice of the compactification.
In the following we shall, therefore, consider cases in which the stability of the Dark Matter particle is not the result of such an ad-hoc symmetry, but rather a residual symmetry naturally present in the model due to its geometry and its compactification.

This idea proves the need to go beyond the minimal UED model in 5 dimensions \cite{Appelquist2000}, where it cannot apply without enforcing parity by hand \cite{Servant2003a}. This has motivated the construction of models in more than 5 dimensions, of which the flat 6-dimensional (6D) UED model based on the real projective plane \cite{Cacciapaglia:2009pa} is the minimal example. Another motivation to look into the case of six dimensional UED models is the number of matter generations required in order to cancel 
the global $SU(2)$ anomaly, which is precisely three \cite{Dobrescu:2001ae}.

To further explore this project, in this paper we study 6D cases with positively curved extra-space, in particular we focus on quotients of the sphere by its finite symmetry groups. 
We will thus systematically study the possible compact spaces, and select the ones that allow for an exact parity, or continuous symmetry, to be left unbroken. On spherical orbifolds, one of the main hurdles for a realistic model is the presence of light (massless) chiral fermions in the spectrum, as spin-1/2 fields typically acquire a mass from the positive curvature. One, therefore, needs to extend the model in order to cancel such mass contribution.
Proposals to this effect have been put forward in the literature, as in \cite{Maru2010,Dohi2010}, making use of an appealing mechanism for radius stabilisation that employs a magnetic monopole of an extra $U(1)$ \cite{RandjbarDaemi:1982hi} also leading to four dimensional chiral fermions. Another possibility for obtaining realistic fermions in the low energy spectrum of sphere-based models is to localise them on a 4-dimensions subspace. A dynamical mechanism has been proposed in \cite{Frere:2003,Frere:2013,Frere:2015} for the localisation of fermions on a sphere inside a vortex, but while this model has many interesting features, Dark Matter is not one of them because the vortex breaks the symmetries of the sphere. As an alternative compatible with Dark Matter, we propose to study the brane-localisation of fermions on the fixed points naturally arising in orbifolds of the sphere. We leave however the requirement that the fields propagating in the bulk corresponds to SM fields, i.e. gauge vectors and the Brout-Englert-Higgs scalar. Models where the whole SM is localised and Dark Matter is provided by additional singlet fields in the bulk has been proposed, for instance in \cite{Winslow:2010nk}.

Before entering into the details of model building and of theoretical and phenomenological constraints, in the following section we shall consider
some general background material. Indeed, while the subject of finite spherical symmetry groups is known in extensive detail and figures 
as part of the basics of crystallography and chemistry \cite{flurry1980symmetry}, to our knowledge there is no systematic study of the geometric properties of the associated 
orbifolds, \textit{i.e.} the quotient spaces of the sphere by these groups, which would be readily available in the standard notation used in particle 
physics for model building. 

We therefore organise the paper with Section 2 dedicated to spherical orbifolds and their applications, where we identify a minimal model to be studied in detail in the following sections. Section 3 
contains the study of the spectra for scalar and gauge fields, where we apply generalised spin-weighted spherical harmonics using the formalism 
of Newman and Penrose, on the sphere and in our minimal orbifold model. Section 4 consists of a theoretical analysis of the properties of Dark Matter in this model. Section 5 deals with the collider limits by direct detection 
of Kaluza-Klein particles. Section 6 introduces the upper limit on the mass from the relic density abundance of the Dark Matter candidate.
Section 7 discusses how the results obtained in the minimal model introduced in the previous sections can be generalised. Finally section 8 
gives the conclusions, which can be drawn from our study.

\section{Spherical orbifolds}
\label{orbifolds}
Obtaining a realistic spectrum for fermions is one of the most challenging aspects of model building in extra dimensions: the SM consists of chiral fermions which are massless in the electroweak preserving vacuum, thus corresponding to a set of Weyl spinors. However, in spaces with more than 4 dimensions, all irreducible spinor representations are built with at least two independent Weyl spinors: the number of Weyl components in the smallest representation is $2^{\frac{D-2}{2}}$ in even dimensions and $2^\frac{D-1}{2}$ in odd dimensions, see for example \cite{RauschdeTraubenberg:2005aa}.
These extra-components manifest themselves as extra fields in the Kaluza-Klein reduction of a theory with bulk fermions, thus providing additional unwanted massless or light states.
A classical way around this problem is the use of orbifolds: one works in a theory where the space is reduced to its quotient by a discrete isometry group, and the fields in the quotient are fields in the original space with a requirement to have particular identification conditions (\textit{orbifolding conditions}) under the isometry group (the \textit{orbifolding group}). These requirements impose that some wavefunctions in the Kaluza-Klein expansion of the fields vanish (the corresponding modes are \textit{orbifolded out} or \textit{projected out}), thus an appropriate choice of isometry group and identification conditions allows the problematic low-lying modes to be projected out -- therefore making the fermion spectrum realistic. This idea was first used in 5 dimensions \cite{Appelquist2000} and was further applied successfully in 6D cases \cite{Dobrescu:2004zi,Cacciapaglia:2009pa}. Positively curved spaces, however, pose a problem that orbifolds alone cannot solve: the Lichnerovicz theorem \cite{berline1992heat} states that in all such geometries, the lowest lying fermion mode has a non-zero mass (thus being a Dirac spinor), which is not compatible with obtaining the Standard Model fields in the lowest levels. 

In this section we present two alternative ways of obtaining massless chiral fermions in the lowest Kaluza-Klein mode, where orbifolds are used in a non-standard way: the Randjbar-Daemi--Salam--Strathdee (RSS) compactification, which has been studied in the literature, and brane-localisation, which we propose as an alternative. We start by reviewing the discrete spherical symmetry groups and the orbifolds they generate, then we show how the RSS compactification provides realistic fermions but also other undesirable features that are hard to circumvent. Finally we argue for an alternative possibility based on brane-localisation, which we will study in detail.

\subsection{Spherical symmetry groups and their associated orbifolds}
The properties of spherical symmetry groups are well known from the mathematical literature \cite{Conway2002,Conway2008}, so we mainly list them for the purpose of fixing our notation. There is, however, to our knowledge, no systematic study of the various possibilities for model building in particle physics: in particular, we will be interested in the residual symmetries of the orbifolded space, which will allow us to determine the optimal candidate models to contain a dark matter candidate.

\subsubsection{The eight families}
There are eight families of spherical symmetry groups:
\begin{itemize}
\item the 3 cyclic groups: 
  \begin{itemize}
   \item $C_n$, generated by a $n$-fold rotational symmetry around the vertical axis,
    \item their ``pyramidal'' variants $C_{nv}$, which include $n$ additional vertical mirror planes (except for $n=2$ where the two planes are identical),
    \item their ``horizontal'' variants $C_{nh}$, which include one horizontal mirror plane;
  \end{itemize}
\item \textit{spiegel} groups $S_{2n}$ of $2n$-fold rotation-inversions around the the vertical axis (note that $S_{2n+1}=C_{(2n+1)v}$);
\item the 3 dihedral groups:
\begin{itemize}
\item  $D_n$, generated by a $\frac{2\pi}{n}$ rotation around the vertical axis and $n$ $\pi$-rotation axes in the horizontal plane,
\item prismatic variants $D_{nh}$ which have an additional horizontal mirror transformation,
\item anti-prismatic variants $D_{nd}$ which have $n$ additional mirror
  planes containing the vertical axis and the line bisecting a pair of adjacent rotation axes;
\end{itemize}
\item polyhedral groups, based on the symmetries of regular polyhedra. 
  \begin{itemize}
  \item the tetrahedral group $T$ and its two extensions $T_d$ with a reflection containing one rotation axis of $T$ and $T_h$ with a reflection in a plane containing two axes of rotations in $T$,
  \item the octohedral group $O$ and its extension $O_h$ which is extended by a reflection around the horizontal plane.
  \end{itemize}
\end{itemize}
Each one of the above isometry groups can be used to define a spherical orbifold. A more explicit description of the groups based on spherical coordinates can be found in Appendix~\ref{app:orbdetails}.

\subsubsection{Symmetries and fixed points}
The 2-sphere covered by the two additional compact spatial dimensions is invariant under the rotation group $O(3)$, before the orbifold projection. It is thus natural to label the KK modes in the expansion in terms of ``angular momentum'' quantum numbers: the total momentum $l$ and the projection along the $z$-axis $m$. We will see in the later sections that this is indeed a convenient choice for the wavefunction basis. 
After the orbifolding, only a subgroup of $O(3)$ remains as the isometry group of the space. Understanding this group is
particularly important for dark matter as it is what provides the quantum numbers preventing its decay.
We found that there are four non-polyhedral group families that have a non-trivial symmetry group. There are three types of residual symmetries: $O(3)$,
$O(2)$ and reflections. $O(3)$ is associated with full angular momentum selection rules, meaning that both $l$ and $m$ are conserved quantum numbers; $O(2)$, corresponding to rotational invariance around one fixed axis of the sphere, keeps only the additive quantum number $m$ conserved; finally reflections with respect to a plane act as parities on the KK modes. The list of symmetries for each class of orbifolds is provided in Table~\ref{tab:orbprop}.

If the orbifolding group acts freely on the sphere (\textit{i.e.} there is no fixed point), the resulting orbifold is a manifold and each point in the orbifold is an orbit with the same number of elements as the group. On the other hand, if there is a non-trivial element of the group that leaves a point fixed, the resulting point in the orbifold can be seen as a singularity\footnote{The canonical example for this behavior is the orbifolding of the plane by the $Z_n$ rotation group around a point x. The resulting quotient space is a cone with a singularity at x. For an example in which a curvature of the form $R_x\delta(x)$ is consistently assigned to such a point, see \cite{Carroll}.}. Note that this means that a point does not need to be stable under the action of the whole group to be singular but we will still use the terms fixed and singular interchangeably in the following. Finally, when the set of singular points form a line on the sphere, we will refer collectively to them as a fixed line and symmetries in the orbifold space might impose some features of the theory to be identical over the whole line, essentially making it a 5D brane.

Once a field theory is defined on the orbifold space, fixed points or lines manifest themselves as branes where localised
operators or fields are allowed. Without imposing additional symmetries, higher order operators appear, thus breaking the $O(3)$ symmetry of the sphere to the unbroken subgroups described above. For the presence of a dark matter candidate, therefore, it is crucial to understand both the presence of fixed points/lines and the residual symmetries.
These special regions, and the corresponding localised operators, will play and important role for model building and will be discussed 
in more detail in the next sections. A summary table of orbifold properties is provided in Table \ref{tab:orbprop}.
\begin{table}[!h]
\begin{center}
\begin{tabular}{|c|c|ll|ll|}
\hline
Group & Symmetry & \multicolumn{2}{c|}{Singular points} & \multicolumn{2}{c|}{Lines}\\ \hline
\multirow{2}{*}{$C_n$} & \multirow{2}{*}{$O(2)_z$} & \multirow{2}{*}{2:} & North pole & \multirow{2}{*}{0} & \\
 & & & South pole & &\\\hline
$C_{nh}$ & $O(2)_z$ & 1: & Poles & 1: & Equator segment  \\\hline
\multirow{2}{*}{$C_{nv}$} & \multirow{2}{*}{None} & \multirow{2}{*}{2:} & North Pole &   \multirow{2}{*}{2:} & Eastern Meridian  \\
& & & South Pole & & Western Meridian  \\\hline
\multirow{3}{*}{$D_{n}$} & \multirow{3}{*}{$M(xz),\ M(xy)$}  & \multirow{3}{*}{3:} &  Poles &  \multirow{3}{*}{0} &   \\ 
& & & East Meridian End & &   \\
& & & West Meridian End & &   \\\hline
\multirow{4}{*}{$D_{nh}$} & \multirow{4}{*}{None}  & \multirow{4}{*}{4:} &  North Pole &  \multirow{4}{*}{0} &   \\ 
& & & Equator Midpoint & &   \\
& & & East Meridian End & &   \\
& & & West Meridian End & &   \\\hline
\multirow{3}{*}{$D_{nv}$} & \multirow{3}{*}{None} & \multirow{3}{*}{1:} & \multirow{3}{*}{Poles} & \multirow{3}{*}{3:} & Eastern Edge   \\
&&&&& Southern Edge\\
&&&&& Western Edge\\\hline
$S_2$ & $O(3)$  & 0 & & 0 & \\\hline
$S_{2n}$, $n>1$ & $O(2)_z$ & 1: & Poles & 0 &  \\\hline
\multirow{3}{*}{$T$}&\multirow{3}{*}{$M(xy)$}& \multirow{3}{*}{3:} & North-South Corners & \multirow{3}{*}{0} &\\
& & & East Corner & &   \\
& & & West Corner & &   \\\hline
\multirow{3}{*}{$T_d$} & \multirow{3}{*}{None} & \multirow{3}{*}{3:} & North Corner & \multirow{3}{*}{1:} & \multirow{3}{*}{Equator segment}\\
& & & East Corner & &   \\
& & & West Corner & &   \\\hline
\multirow{2}{*}{$T_h$} & \multirow{2}{*}{None} &\multirow{2}{*}{2:} & North-South Corners & \multirow{2}{*}{1:} & \multirow{2}{*}{Western Edge}\\
& & & East Corner & &   \\\hline
\multirow{3}{*}{$O$}&\multirow{3}{*}{$M(xy)$}& \multirow{3}{*}{3:} & North Corner & \multirow{3}{*}{0} &\\
& & & East-West Corners & &   \\
& & & Equator Midpoint & &   \\\hline
\multirow{3}{*}{$O_h$}&\multirow{3}{*}{None}& \multirow{3}{*}{3:} & North Corner & \multirow{3}{*}{1:} & \multirow{3}{*}{Equator segment}\\
& & & East Corner & &   \\
& & & West Corner & &   \\\hline
\end{tabular}
\end{center}
\caption{Properties of the spherical orbifolds. }
\label{tab:orbprop}
\end{table}

\subsection{The RSS mechanism and its phenomenological applications}

In six dimensions, a Dirac fermion contains four Weyl components, which can be grouped into two irreducible 6D--chiral representations:

\begin{equation}
  \label{eq:1}
  \Psi_{6D} = \Psi_+ + \Psi_- = \begin{pmatrix} \chi_+ \\ \bar\eta_+ \\ 0 \\ 0\end{pmatrix} + \begin{pmatrix} 0 \\ 0 \\ \chi_- \\ \eta_- \end{pmatrix}\,.
\end{equation}

We can write the Lagrangian for a chiral 6D fermion $\Psi_+$, in terms of its components, in the following way:
\begin{equation}
  \label{eq:3}
  {\cal L}_{\Psi_+} = \int \sqrt{-g} dx^6\  i(\chi_+^\dagger)\sigma_\mu \partial^\mu \chi_+ +  i(\bar\eta_+^\dagger)\sigma_\mu \partial^\mu \bar\eta_+i+\left( \chi^\dagger\bar\Omega_+ \bar\eta_+ + \bar\eta_+^\dagger \Omega_+ \chi_+  \right)\,,
\end{equation}
where $\Omega_+$ and $\bar \Omega_+$ are two differential operators which have conjugate eigenbases and opposite eigenvalues. 
As such operators connect $\chi_+$ and $\bar \eta_+$, their KK expansion contains the same physical states and $\Omega_+$ and $\bar\Omega_+$ provide mass terms which connect the two chiralities in each level thus making explicit that we have vector-like fermion levels. The problem with fermions on the sphere is that $\Omega_+$  does not have a null eigenvalue, thus no massless mode is present in the spectrum.

A significant change to this picture can be brought by adding a gauge field background, which will generate terms of the same kind as $\Omega_+$ since both are connexions. In particular, if a $U(1)$ gauge field $\v A$, with coupling $g$, has a magnetic monopole background on the sphere
\begin{equation}
  \label{eq:6}
  \v A = \frac{n}{2g} \cos \theta d\phi,
\end{equation}
the connexion operators are changed in a way that allows chiral fermions in the lowest mode. Indeed, the monopole gauge background provides a 
$g$-independent term in the Dirac equations and breaks the symmetry in the expansions of $\chi$ and $\bar\eta$. Depending on the sign of the charge, 
either one has a constant wavefunction among its modes while the other only has non-zero kinetic energy modes. As a result, there is only one massless 
Weyl fermion KK mode for each 6D chiral fermion which interacts with the monopole background.

Besides the obvious motivation from model building, there are some stronger arguments for introducing this gauge background from the study of the Einstein 
equations on the background $\mathbb{M}_4\times S^2$. In this case, it was established by Randjbar-Daemi, Salam and Strathdee \cite{RandjbarDaemi:1982hi}
that having a stable geometry is not possible with an empty spacetime, but that a monopole gauge background fixes the sphere radius, depending on the 
gauge coupling:
\begin{equation}
  \label{eq:8}
  R = \frac{n\sqrt{8 \pi G_6}}{2g}
\end{equation}
where $G_6$ is the 6-dimensional gravitational constant, related to the usual Newton constant by $\frac{G_6}{R^4}=\frac{G}{R^2}$.
The effective 4D gauge coupling for this gauge field therefore scales like $\sqrt{G}/R$ and numerically is of order $10^{-8}$ for\tev{} extra-dimensions. While this is a very small coupling that protects the theory from any observable effects at colliders, it has been pointed out \cite{Dohi2010} that measurements of short range gravitational interactions can put bounds on such a gauge theory, while even a small mass around a few$\unit{}{\electronvolt}$ is sufficient to hide the new gauge boson from these experimental searches \cite{Geraci2008}.
Two models have been proposed in the past that make use of the RSS mechanism to have chiral fermions in their low-energy spectra. They both rely on some form of explicit breaking of the $U(1)$ symmetry. 

The first model \cite{Maru2010} is defined on the orbifold called $S^2/C_2$ in our notation. Their proposal for breaking the gauge field introduced in their model is to use the gauge anomaly present in their theory if only the Standard Model fermions are included. As a result, an ${\cal O}(1/R)$ bulk mass is expected to be given to the gauge field, putting the new interaction far beyond any short range gravity test. The situation is similar in \cite{Dohi2010}, where the gauge symmetry is broken by the orbifold identification conditions: by imposing a charge conjugation violating identification in the geometry $G_2$, the zero-mode of the gauge field is eliminated from the spectrum, leaving only massive, ${\cal O}(1/R)$ excitations. In both cases, gauge invariance is explicitly broken. Besides aesthetics, we think that this is potentially troublesome: the gauge breaking allows for a large number of unconstrained bulk operators with power-law scaling -- potentially largely changing the tree-level picture that the authors studied. In particular, new quadratic operators are allowed -- first of which the mass term -- modifying the equations of motions and therefore spoiling the GR argument due to the large corrections these operators introduce into the original Einstein-Maxwell equations. It is not at all clear that a stable radius is still possible under these conditions. While it is definitely true that most UED models do not provide spaces that are solutions to gravitational equations -- which is justified by their being effective field theories only valid up to at best tens of\tev{} -- this observation does take away a lot of the naturalness that the two models seem to have at tree level.

As an alternative to the extra complications that the RSS mechanism brings to particle models, we propose instead to localise fermions on the fixed points of spherical orbifolds. Such a construction does not provide a compactification mechanism and leaves gravity out of the picture, but has a reasonable justification: since localised operators are naturally generated on the fixed points of orbifolds, it is not far-fetched to allow for some fields to be localised as well.

\subsection{Selecting an orbifold for a dark matter model}

We wish to focus our attention on the simplest geometry compatible with the existence of dark matter and massless chiral fermions. 
The first requirement is obviously that a symmetry exists after orbifolding, so that some KK modes will be stablilised. This leaves only four 
possible candidate families: $C_n$, $C_{nh}$, $D_n$ and $S_{2n}$. We also wish to localise fermions on fixed points, as this allows avoiding 
the usual problems of bulk fermions requiring a special treatment in order to let survive only a chiral 4D fermion from a 6D one, which further 
excludes $S_2$. Among the remaining possibilities,  $S_{2n}$ with $n>1$ provides the simplest framework for this localisation 
since it has exactly one fixed point, leaving no ambiguity as to where the fermions should be localised. The class $C_{nh}$ also has a single fixed point with the addition of a fixed line: the line is equivalent to a circle, thus localising a fermion field there cannot give rise to chiral massless modes. Other cases have more than one fixed point: 2 for $C_n$ and $C_{nv}$, and 4 for $D_n$.
Note that the two simplest polyhedral orbifolds, $T$ and $O$, also have a left-over parity, however determining the spectra is mathematically not an easy task.
In the following sections, we will therefore study 
the phenomenology of the model with geometry $\mathbb{M}_4\times S^2/S_4$ in detail, and then generalise the results obtained to the other 
possible spaces.

\section{Spectra on $S^2$ and $S^2/S_4$}
\label{spectra}

Starting from a spherical geometry, it is not surprising that angular momentum plays a major role in the classification of the wave-functions of KK modes, including the case of orbifolds. Spin-weighted spherical harmonics provide a 
truly natural framework for the study of arbitrary-spin equations of motion on spaces with spherical symmetry. They were introduced 
in \cite{Gelfand1958} for the study of representations of the Lorentz group, then independently rediscovered in \cite{Newman1966}, 
and subsequently used to study gravitational radiation \cite{Penrose1984}. An introductory review can be found in \cite{Castillo2007}. 
In recent years, spin-weighted harmonics have been successfully applied to finding compact solutions to the equations of motion of 
scalars, gauge bosons and spin-\half fermions for the derivation of Kaluza-Klein spectra \cite{Dohi2010, Dohi2014}. While previous solutions 
existed (see \cite{Camporesi1996, Abrikosov2002, Maru2010} and others), the simplicity of the derivation shows that the spin-weighted
spherical-harmonic expansion is the best suited formulation of these results.

In this section, we restate basic results about spin--weighted harmonics, which will be used to obtain physical degrees of freedom of
the Kaluza-Klein expansion, following the derivation of \cite{Dohi2014}.

\begin{dfn}[Spin Weight]
Let $f$ be a function defined as
\begin{equation}
  \label{eq:1}
f:  \left\{ \begin{array}{ccc} 
S^2\times S^1 &\rightarrow& \mathbb{C}\\
(\Omega,\uv{v}) &\mapsto & f(\Omega, \uv{v})
\end{array}\right.
\end{equation}
where $\uv{v}$ defines a right-handed basis on the tangent space of
$S^2$ at the point $\Omega$. The function $f$ has spin-weight $s$ if
$f(\Omega,e^{i\theta} \uv{v})=e^{is\theta}f(\Omega,\uv{v})$
\end{dfn}

These function are spin-$s$ representations of the $U(1)$ symmetry of the tangent space of $S^2$. As for usual spin representations, 
they can be built from products of the spin-\half representations, as illustrated explicitly in \cite{Castillo2007}. The spin-weighted
spherical harmonics $\phantom{a\hspace{-0.5em}}_{s\hspace{-0.1em}}Y_{lm}$ provide a particularly nice basis for functions of
a given spin weight and in particular, the usual spherical harmonics are the trivial representation, generated
from the antisymmetric product of the spin-\half basis. The details for this construction being of no direct consequence for our
discussion, which only uses the algebraic properties of the spin-weighted spherical harmonics, we refer the readers to
\cite{Castillo2007} for further details and derivations.
\begin{dfn}[Newman-Penrose operator] We define the operator $\eth$, and its conjugate $\bar\eth$, as
\begin{equation}
  \label{eq:2}
    \begin{array}{rclrcl}
     \phantom{a\hspace{-0.5em}}_{s\hspace{-0.1em}}\eth &=& - \left( \partial_\theta + i \frac{\partial_\phi}{\sin\theta}
  - s \cot \theta\right),\ \  &      \phantom{a\hspace{-0.5em}}_{s\hspace{-0.1em}}\bar\eth &=& - \left( \partial_\theta - i \frac{\partial_\phi}{\sin\theta}
  + s \cot \theta\right).
\end{array}
\end{equation}
\end{dfn}
These operators appear naturally in calculations on the sphere.
We will often omit the index $s$ when it matches the explicit spin-weight index of the function we apply it to.
These operators respectively increase and decrease the spin-weight of spherical harmonics:
\begin{equation}
  \label{eq:3}
  \begin{array}{rclrcl}
    \eth \phantom{a\hspace{-0.5em}}_{s\hspace{-0.1em}} Y_{lm} & = &
\sqrt{l(l+1)-s(s+1)} \phantom{a\hspace{-0.2em}}_{s+1\hspace{-0.1em}}
Y_{lm},\ \  & \bar\eth \phantom{a\hspace{-0.5em}}_{s\hspace{-0.1em}} Y_{lm} & = &
-\sqrt{l(l+1)-s(s-1)} \phantom{a\hspace{-0.2em}}_{s-1\hspace{-0.1em}}
Y_{lm} \,.
  \end{array}
\end{equation}
The spin-weighted spherical harmonics are eigenfunctions of the
related operators
\begin{equation}
  \label{eq:4}
  \begin{array}{rclrcl}
    \eth\bar\eth \phantom{a\hspace{-0.5em}}_{s\hspace{-0.1em}} Y_{lm} & = &-
(l(l+1)-s(s-1)) \phantom{a\hspace{-0.2em}}_{s\hspace{-0.1em}}
Y_{lm},\ \  & \bar\eth\eth \phantom{a\hspace{-0.5em}}_{s\hspace{-0.1em}} Y_{lm} & = &
-(l(l+1)-s(s+1)) \phantom{a\hspace{-0.2em}}_{s\hspace{-0.1em}}
Y_{lm}  \,,
  \end{array}
\end{equation}
which reduce to the Laplace-Beltrami operator for $s=0$.

We will also need to use the symmetry properties of these functions:
\begin{equation}
  \begin{array}{c}
    \phantom{a\hspace{-0.2em}}_{s\hspace{-0.1em}}
Y_{lm}^*=(-1)^{m+s}\phantom{a\hspace{-0.2em}}_{s\hspace{-0.1em}}
Y_{lm}\\
\phantom{a\hspace{-0.2em}}_{s\hspace{-0.1em}}
Y_{lm}(\pi-\theta,\phi+\pi) = (-1)^j \phantom{a\hspace{-0.2em}}_{-s\hspace{-0.1em}}
Y_{lm}(\theta,\phi)
  \end{array}
\end{equation}
as well as the overlap integrals that will determine the selection
rules between Kaluza-Klein levels:
\begin{equation}
  \label{eq:5}
  \begin{array}{rll}
    \int d\Omega \phantom{a\hspace{-0.2em}}_{s\hspace{-0.1em}}
Y_{lm}^* \phantom{a\hspace{-0.2em}}_{s\hspace{-0.1em}}
Y_{l'm'} =& \delta_{ll'} \delta_{mm'}&\\
   \int d \Omega \phantom{a\hspace{-0.2em}}_{s_1\hspace{-0.1em}}Y_{l_1m_1}\times \phantom{a\hspace{-0.2em}}_{s_2\hspace{-0.1em}}Y_{l_2m_2}\times\phantom{a\hspace{-0.2em}}_{s_3\hspace{-0.1em}}
Y_{l_3m_3} =&& \\
 \sqrt{\frac{l_1(l_1+1)l_2(l_2+1)l_3(l_3+1)} {4\pi}}&\left(\begin{array}{ccc}
  l_1 & l_2 & l_3\\
m_1 & m_2 & m_3
\end{array}\right)
\left(
\begin{array}{ccc}
  l_1 & l_2 & l_3\\
-s_1 & -s_2 & -s_3
\end{array}\right)&
  \end{array}
\end{equation}
Using these definitions and properties, we will review the spectra
derived in \cite{Dohi2014} for fields defined on a $\mathbb{M}_4\times S^2$.

\subsection{Scalar bosons on a sphere}

Scalar bosons have the most straightforward expansion as their quadratic Lagrangian is expressed with a Laplace--Beltrami
operator. Let us look at the reduction for a scalar field $\Phi(x^\mu,\theta,\phi)$ with 6D Lagrangian
\begin{equation}
  \label{eq:6}
  {\cal L}_\Phi=\Phi\partial_\mu \partial^\mu \Phi + \Phi\Delta_{S^2}\Phi-m_\Phi^2 \Phi\Phi\,.
\end{equation}
The Laplace--Beltrami operator is diagonal in the spherical harmonics basis, so we can expand
\begin{equation}
  \label{eq:7}
  \Phi(x^\mu,\theta,\phi)=\sum_{lm} \frac{\Phi_{lm}(x^\mu) }{R}Y_{lm}(\theta,\phi)\,,
\end{equation}
which allows to write the 4D-reduced Lagrangian
\begin{equation}
  \label{eq:8}
  \begin{array}{rl}
    {\cal L}_{\Phi 4D} &= R^2\int d\Omega {\cal L}_\Phi\\
&= \displaystyle\sum_{lm} \Phi_{lm}^\dagger\partial_\mu \partial^\mu \Phi_{lm} -
\left(m_\Phi^2+l(l+1)\right) \Phi^\dagger_{lm} \Phi_{lm}\,.
  \end{array}
\end{equation}

\subsection{Gauge bosons on a sphere}
The description of the properties of spin-1 gauge bosons can be obtained in a way, which is very similar to the derivation for the scalars.
At the same time some differences are crucial in terms of applications to phenomenology and model building, as discussed in the following.

\subsubsection{Degrees of freedom}
As usual for gauge bosons in extra-dimensional theories, the gauge field $A_M$ breaks into two parts: a 4D vector spanning the 4 components 
pointing into ordinary space-time directions, and 4D scalar fields pointing along the extra space. In the KK expansion, the scalar fields contain both Goldstone bosons for the massive vectors
in the Kaluza-Klein expansion, and physical scalars. 
The connection between the components $A_{\theta,\phi}$ and the physical degrees of freedom, however, requires a non-trivial redefinition of the fields, unlike in the flat case.
We use the following redefinitions \cite{Dohi2014}:
\begin{align}
\label{eq:9}
\tilde A_{\theta} & = \frac{A_{\theta}}{R}\,, & \tilde{A}_\phi & =  \frac{A_\phi}{R\sin\theta}\,,  & & \nonumber \\
A_\pm &= \frac{1}{\sqrt{2}}(\tilde A_\theta \pm i\tilde{A_\phi})\,, & \tilde A_\theta
&=  \frac{1}{\sqrt{2}}(A_+ + A_-)\,, & \tilde A_\phi&
=  \frac{1}{\sqrt{2}i}(A_+ - A_-)\,, \nonumber \\
A_+&=A_-^\dagger\,, & A_+ &= -i \eth \Xi^\dagger\,, & \Xi &= \frac{1}{\sqrt{2}}(\Phi + i\Theta) \,.
\end{align}
As we shall see from the Lagrangian, $\Phi$ is the physical scalar that will remain in the
spectrum, while $\Theta$ is a Goldstone boson, whose KK levels will be
eaten by the corresponding levels of the vector to give them mass. 

\subsubsection{Gauge fixing}

As in standard gauge theories, we need a gauge fixing part for the
Lagrangian. The usual Feynman-t'Hooft part is present to fix the gauge
of the 4D massless vector, while the rest is determined to eliminate
the mixing between the vector and scalar parts in the quadratic
Lagrangian \cite{Maru2010, Dohi2014}:
\begin{equation}
  \label{eq:10}
  {\cal L}_{gf}=\frac{1}{\xi}tr(f^2)
\end{equation}
where
\begin{equation}
  \label{eq:11}
  f=\eta^{\mu\nu} \partial_\mu A^\nu +
  \frac{\xi}{R^2\sin\theta}\partial_\theta \sin\theta A_\theta +
  \frac{\xi}{R^2\sin^2\theta}\partial_\phi A_\phi\,.
\end{equation}

\subsubsection{Equations of motion and expansion}

Using these definitions, we can write the quadratic 6D Lagrangian for
the vector component:
\begin{equation}
  \label{eq:12}
  {\cal L}_{A^\mu}=A_\mu \left[ \eta^{\mu\nu} \left(\Box +
    \frac{1}{R^2}\phantom{a\hspace{-0.5em}}_{1\hspace{-0.1em}}\bar\eth \phantom{a\hspace{-0.5em}}_{0\hspace{-0.1em}}\eth\right) - \left(1-\frac{1}{\xi}\right)\partial^\mu\partial^\nu\right]A_\nu
\end{equation}
and for the scalars:
\begin{equation}
  \label{spectra:eq:13}
  {\cal L}_{\Phi,\Theta} = \Phi\;
  \phantom{a\hspace{-0.5em}}_{1\hspace{-0.1em}}\bar\eth
  \phantom{a\hspace{-0.5em}}_{0\hspace{-0.1em}}\eth \left( \Box -
    \frac{1}{R^2}\phantom{a\hspace{-0.5em}}_{1\hspace{-0.1em}}\bar\eth
    \phantom{a\hspace{-0.5em}}_{0\hspace{-0.1em}}\eth\right)\Phi +
  \Theta\; \phantom{a\hspace{-0.5em}}_{1\hspace{-0.1em}}\bar\eth
  \phantom{a\hspace{-0.5em}}_{0\hspace{-0.1em}}\eth \left( \Box - \frac{\xi}{R^2}\phantom{a\hspace{-0.5em}}_{1\hspace{-0.1em}}\bar\eth \phantom{a\hspace{-0.5em}}_{0\hspace{-0.1em}}\eth\right)\Theta \; ;
\end{equation}
thus confirming that $\Theta$ is a Goldstone boson. The Penrose-Newman
operators appearing in this decomposition show that the expansion for
both types of fields has to be in terms of usual spherical harmonics.
\begin{equation}
  \label{eq:14}
  \begin{array}{rl}
    A_\mu &=\displaystyle \sum_{lm} \frac{A_{\mu lm}}{R} Y_{lm}\,,\\
    \Phi &=\displaystyle \sum_{l>0,m} \frac{\Phi_{ lm}}{R\sqrt{l(l+1)}} Y_{lm}\,,\\
    \Theta&= \displaystyle\sum_{l>0,m} \frac{\Theta_{lm}}{R\sqrt{l(l+1)}} Y_{lm}\,.\\
  \end{array}
\end{equation}
There is no zero mode for the scalar due to the extra operator in the Lagrangian. For both physical fields, the masses of level $(l,m)$ are equal to $\sqrt{l(l+1)}$ in units of the radius $R$.

\subsection{Spectra on $S^2/S_4$}

\begin{figure}[!tb]
  \centering
\includegraphics[width=0.7\textwidth]{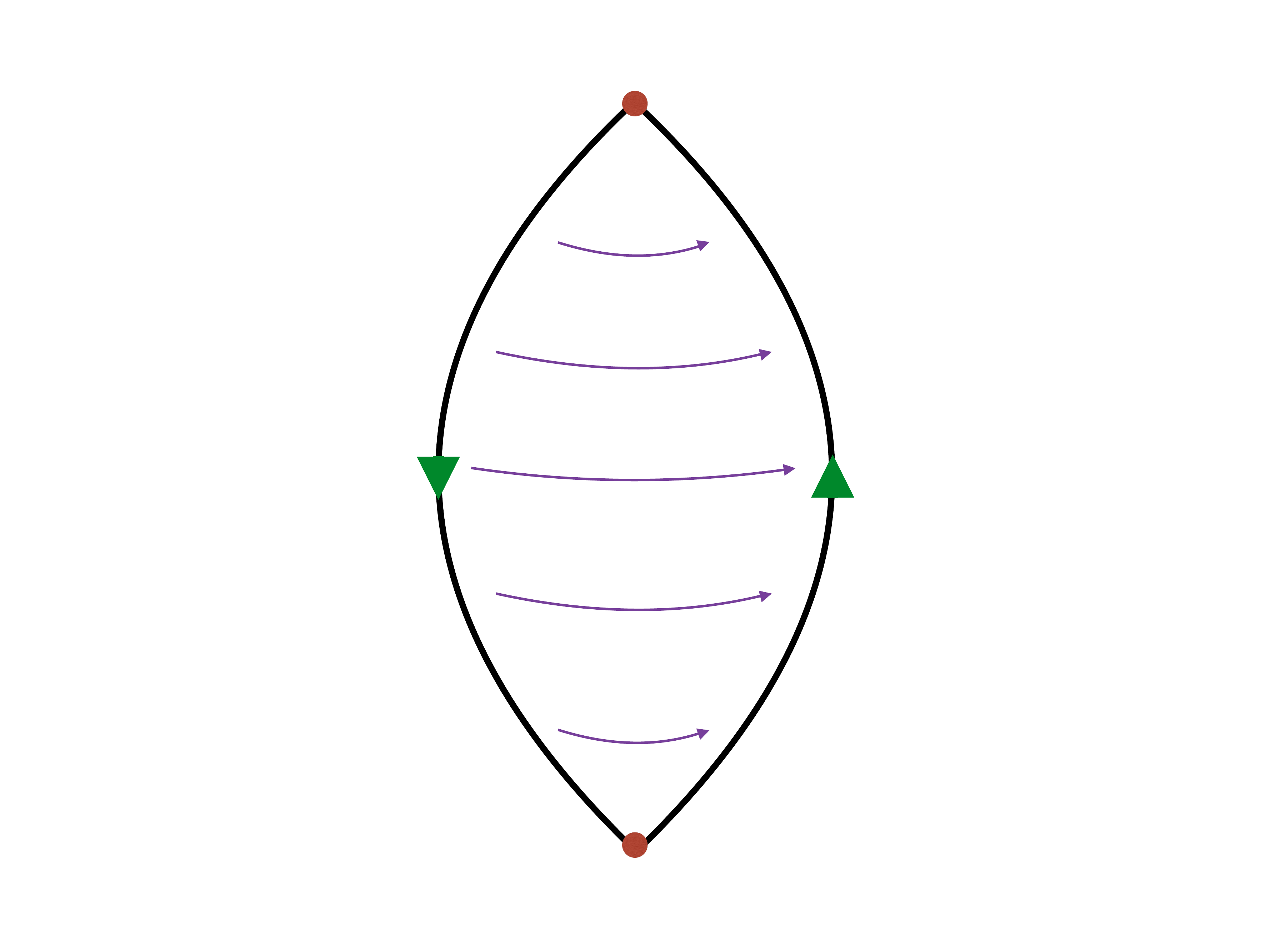}
  \caption{Representation of $S^2/S_4$ as a fundamental domain on
    $S^2$ with boundary conditions in green - using standard
    nomenclature for $\mathbb{R}^2$ orbifolds \cite{Nilse2006, Cacciapaglia:2009pa}. There is one singular point: the orbit of the North and South poles of the sphere, which are identified.
\ This figure helps illustrate that the residual $U(1)$ symmetry of the
  orbifold is in fact $\phi$ translational invariance, which is
  suggested in purple arrows}
\label{fig:S2S4rot}
\end{figure}

Having summarised the KK decomposition on a sphere, we now turn to the study of how the spectrum is modified on the orbifold $S^2/S_{2n}$.
This class of orbifolds has a single generator, and the projection consists in identifying the points $(\theta, \phi) \sim G (\theta, \phi)$.
The orbifolding group generator is the following
\begin{equation}
G\left\{
\begin{aligned}
\theta&\rightarrow\pi-\theta\\
\phi&\rightarrow\phi+\pi\left(1+\frac{1}{n}\right) 
\end{aligned}
\right.
\end{equation}
and one can define the fundamental domain as shown in Figure \ref{fig:S2S4rot}.
Then we can easily figure out the behaviour of $Y_{lm}$ under $G$ as
$Y_{lm}$ has parity $l$ and depends on $\phi$ as $e^{im\phi}$: thus we have
\begin{equation}
Y_{lm}(G\Omega) = (-1)^lY_{lm}(\Omega)e^{im\pi/n}\,.
\end{equation}
For a field $\Psi$, the orbifolding condition can be written in
general as $G_\Psi\Psi(x,G\Omega)=\Psi(x,\Omega)$ where
$G_\Psi$ is the representation of $G$ acting on $\Psi$. If we impose
that the representation of $G$ is trivial for 6D scalars and gauge
vectors, i.e. $G_\Psi = 1$, then the modes $(l,m)$ that survive the orbifolding are:
\begin{itemize}
\item $l$ even: $\frac{m}{n}$ is an even integer so $m=0,2n,4n,...$
\item $l$ odd: $\frac{m}{n}$ is an odd integer so $m=n,3n,...$
\end{itemize}
Note that a zero mode $(l,m) = (0,0)$ is consistent. This expansion is valid for the scalar (Higgs), and for spin-1 states.
For the scalar components of gauge vectors, the constraint that the field transforms as a 6D vector, imposes that the transformation of the scalar components are
$G_A A_\theta = - A_\theta$ and $G_A A_\phi=A_\phi$, which trickles
down to imposing that $\Phi$ has the following expansion:
\begin{itemize}
\item $l$ even: $\frac{m}{n}$ is an odd integer so $m=n,3n,...$ 
\item $l$ odd: $\frac{m}{n}$ is an even integer so $m=0,2n,4n,...$
\end{itemize}
In the case $n=2$, which we will study in detail, this means that the first level $l=1$ only contains gauge scalars with $m=0$ and the next level is $l=2$, with gauge vectors and scalar fields at $m=0$, and gauge scalars at $m=\pm 2$.

\section{The lightest rotating particle}
\label{dmlevel}


As we have stated in Section \ref{orbifolds}, we are interested in the
orbifold $S^2/S_4$ for our model because of its geometric
features. Indeed, we found that besides its single fixed point, there
is a residual global $U(1)$ symmetry of this space. Due to Noether's Theorem,
this symmetry is associated with a quantum number that is conserved,
which in turn means that the lightest state with the lowest non-zero value of
this quantum number will be stable.
As we show in Figure \ref{fig:S2S4rot}, the residual symmetry, expressed
in usual spherical coordinates, is simply $\phi$ translational
invariance, which is directly associated to the angular momentum
quantum number $m$. The other quantum number $l$, is not conserved.

\paragraph{How $l$ non-conservation appears:} the breaking of the full
angular momentum conservation on the sphese is not completely obvious. Indeed, at
tree level, due to the expansion in spherical harmonics, there seems
to be the same selection rules as in the full $S^2$ case. At one loop
however, the renormalisation of the theory requires localised
operators to appear as counter--terms on the fixed point, which will affect Kaluza-Klein
modes in a $l$-dependent way and break $l$-selection rules.
Additional terms that do not conserve $l$ are due to non-local, finite, loop effects
generated by non-contractible orbits on the orbifold.

\paragraph{The Dark Matter level:} we have shown that the lightest
state with non-zero $m$ is automatically stable from Noether's
theorem. Therefore, from the discussion of Section \ref{spectra}, the
LRP (Lightest Rotating Particle, i.e. our Dark Matter candidate) will be a scalar gauge boson in level $(2,\pm 2)$. In the next
sections, we will show that, to a good approximation, the lightest
state is the scalar excitation of the photon field, which is consistent with
the basic expectation of a neutral dark matter candidate.

\paragraph{Bulk fields:} to obtain a realistic model, only the electroweak gauge bosons and the scalar field inducing electroweak symmetry breaking will be allowed to propagate in the bulk. Fermions are localised on the fixed points in order to be chiral, while the gluon field is localise to avoid the presence of a stable coloured scalar in the spectrum. In fact, the only sector that connects gluon fields to the electroweak bosons are quarks: being localised, quark do not couple to the scalars in the DM level, and thus they cannot induce decays of the scalar gluon to the Dark Matter candidate.

\subsection{Quadratic Lagrangian for level $(2,\pm 2)$ gauge scalars}


The levels $(2, \pm2)$ in our model only contain scalar gauge fields. In the KK
expansion, one would therefore expect one scalar state in each level, and with equal mass.
In fact, the two levels $(2,\pm 2)$ are related by charge conjugation: since
$Y_{2,2}^\dagger = Y_{2,-2}$, we also have $\Phi_{2,2}^\dagger=\Phi_{2,-2}$. 
In other words, for each generator of the gauge group in the bulk, there are 2 real scalars
which form a complex scalar charged under the global $U(1)$. The global symmetry, therefore, acts as a rotation on the two degenerate levels ($O(2)$). This is in fact true for any tier with non-vanishing $m$.

The physical part of a gauge scalar's
kinetic lagrangian is given in Equation \ref{spectra:eq:13}:
\begin{equation}
\begin{array}{lll}
\mathcal{L} &= & \int R^2 d\Omega \ \Phi\; \hspace{0.01em}_1\bar\eth_0\eth \left( \Box -
  \frac{\hspace{0.01em}_1\bar\eth_0\eth}{R^2}\right) \Phi \ \ni\  + \Phi_{2,-2}\left( \Box - \frac{6}{R^2} \right) \Phi_{2,2} + \Phi_{2,2}\left( \Box - \frac{6}{R^2} \right) \Phi_{2,-2} \\
 & = &  \Phi_{2,2}^\dagger\left( \Box - \frac{6}{R^2} \right) \Phi_{2,2} +
 \Phi_{2,2}\left( \Box - \frac{6}{R^2} \right) \Phi_{2,2}^\dagger \ =\
 2 \Phi_{2,2}^\dagger \left( \Box - \frac{6}{R^2} \right)
\Phi_{2,2} \mathrm{\ \ }\\
&= & \varphi^\dagger \left( \Box - \frac{6}{R^2} \right)  \varphi\\
\end{array}
\end{equation}
Where $\Phi_{2,2}=\frac{\varphi}{\sqrt{2}}$ is the physical definition
for a gauge scalar field in level $(2,\pm 2)$ without mixing or
symmetry breaking.

The symmetries of the model also allow to write localised interactions on the fixed point: as we have argued above, such terms do not break the $U(1)$ protecting the Dark Matter candidate. Furthermore, localised operators are always required as counter--terms of the divergent 6D loops, thus they are necessary for a correct renormalisation of the theory.

\subsection{Mass splittings: electroweak and loop effects}

Now that we have singled out which Kaluza-Klein tier will contain the dark matter candidate of our model, we need to find which field in this tier is actually the lightest, and will hence be stable. There are two sources of mass splitting for same-tier Kaluza-Klein particles: loop effects, which are usually dominant, and electroweak effects from the Higgs sector, which are most often negligible. In our case, however, the situation may be reversed as the dominant log-divergent contribution to the loop vanishes and one is left with subleading finite contributions.

\subsubsection{Electroweak symmetry breaking: tree-level spectrum}
At tree level, the only source of mass splitting among the $U(1)_Y$
and $SU(2)_L$ fields is the Higgs mechanism. Let us first work out the electroweak symmetry breaking mechanism in this geometry.

\paragraph{Electroweak symmetry breaking.}
The 6D Higgs Lagrangian is defined as 
\begin{equation}
  \label{eq:10}
  {\cal L}_H^{6D} = (D_M\phi)^\dagger (D^M\phi) +\mu_{6D}^2 \phi^\dagger \phi - \lambda_{6D}
  (\phi^\dagger \phi)^2
\end{equation}
where $\phi$ is a $SU(2)$ doublet:
\begin{equation}
  \label{eq:15}
  \phi = \frac{1}{\sqrt{2}}\left( 
    \begin{array}{c}
      \phi_1+i\phi_2\\
      v_{6D} + H + i\phi_3
    \end{array}
\right)\,.
\end{equation}
The Kaluza-Klein expansion is formulated in terms of spherical harmonics:
\begin{equation}
  \label{eq:11}
  \phi(x^\mu,\Omega) = \sum \frac{Y_{lm}(\Omega) }{R} \phi_{lm}(x^\mu)\,.
\end{equation}

The vacuum expectation value of the field in the minimum energy state has to have a constant wavefunction to minimise kinetic energy, so its value is determined by the
minimum of the potential as: 
\begin{equation}
  \label{eq:12}
v_{6D} = \sqrt{\frac{\mu^2_{6D}}{2\lambda_{6D}}}=\frac{v_{4D}}{\sqrt{4\pi}R}\,.
\end{equation}
Because the situation is the same as in the Standard Model in the lowest Kaluza-Klein mode $(0,0)$, we know what to expect in the Unitary gauge: the Higgs doublet will only contain one physical state, $H_{00}$, which is the Standard Model Higgs boson. In the higher modes, however, the situation is more complex as there are already Goldstone bosons for the excitations of the $W^\pm$ and $Z$ bosons, coming from the gauge scalar levels. Counting degrees of freedom, it appears that there should be 4 physical scalar excitations per KK level which are a mixture of the Higgs and scalar gauge fields, and we can find their exact content by looking at the gauge-fixing Lagrangian to find mass eigenstates, as was done in \cite{Cacciapaglia:2009pa}. 

The gauge-fixing Lagrangian for the broken $SU(2)\times U(1)$ is obtained by expanding on the pure Yang-Mills gauge-fixing term with the requirement that the quadratic terms mixing the Higgs field and gauge bosons be cancelled, yielding the expression:
\begin{align}
{\cal L}_{gf}=&-\frac{1}{\xi} tr(f^a f_a)\\
f^a=& \eta^{\mu\nu} \partial_\mu A_\nu^a + \frac{\xi}{R^2 \sin \theta } \partial_\theta \left( \sin\theta A_\theta^a \right) + \frac{\xi}{R^2 \sin\theta}\partial_\phi A_\phi^a \\
&-\xi gF_i^a \phi_i
\end{align}
where $a\in \left\{ 1,2,3,Y\right\}$ and $gF_i^a=g_{(a)}F_{ij}^a\avg{\phi}^j$ is the the same matrix as in the typical $R_\xi$ gauge-fixing in broken 4D gauge theories. The resulting Lagrangian terms can be sorted by their order in $\xi$

\begin{itemize}
\item ${\cal O}(\xi^{-1})$: correction to the vector propagators that vanish in the unitary limit
\item ${\cal O}(\xi^{0})$: terms that kill mixed quadratic terms between the Higgs, gauge vectors and gauge scalars
\item ${\cal O}(\xi^{1})$: $\xi$-dependent mass matrix that fixes which combination of the Higgs components and gauge scalars are Goldstone bosons and which are physical fields.
\end{itemize}

This matrix must be taken KK-level by KK-level and yields $l$-dependent admixtures of the two components in the physical state, which is its null vector. To give a simplified picture of this mixing, we show the mixing matrix in the case of an Abelian Higgs mechanism:
\begin{equation}
M^2_{gf}=\frac{\xi}{R^2}\begin{pmatrix}g^2 \zeta^2 & g\zeta \sqrt{l\left(l+1\right)} \\  g\zeta \sqrt{l\left(l+1\right)} & l\left(l+1\right)
\end{pmatrix}
\end{equation}
where $\zeta=vR$.
This matrix acts on the vector $(\phi_0,\Theta)^T$ where the first entry is the Higgs component whose zero mode is a Goldstone boson and the second entry is the imaginary part of the canonically normalised gauge scalar $\Xi$.

One can note that $M^2_{gf}$ is of rank 1, showing indeed that one of its eigenvector is integrated out in the Unitary limit $\xi\rightarrow\infty$ while the other does not get a $\xi$-dependent mass and therefore is physical. A null vector of this matrix, corresponding to the physical state is 
\begin{equation}
\begin{pmatrix} 1 \\ -\frac{g\zeta}{\sqrt{l\left(l+1\right)}} \end{pmatrix}\,.
\end{equation}
The Higgs component in the Goldstone boson goes to zero at large $l$, which is expected due to the fact that the mass of the associated vector becomes very large compared to the Higgs vacuum expectation value. 
In particular, already at $l=2$, which is the level of interest for this paper, the mixing is only 10\% , for $1/R\simeq v$ and $g=e$, which justifies treating $\Theta$ completely as a Goldstone boson and keeping the Higgs components as physical.

As a result, the Higgs field has a zero mode behaving like the Standard Model Higgs field, with a constant expectation value and a neutral physical field. In the excited levels, besides the neutral field, there will be an extra neutral scalar and a charged scalar, which have the couplings of the Higgs extra-components since they have a negligible gauge-scalar component. 
Their mass is however a free parameter: in fact, one can add a localised mass term for the scalar field and the way it affects the zero mode and the massive modes is different.
In other words, the mass of the KK excitations of the Higgs receive an additional additive contribution which is unrelated to the mass of the SM-like Higgs boson.
We will discuss the phenomenology of these states in the following sections and now turn to the mass contribution of the vacuum expectation value to the gauge boson masses.

\paragraph{Electroweak mass contribution to the gauge bosons.}
As we show in the Appendix \ref{app:intlagrangian}, the quartic coupling of the Higgs field to
the $SU(2)$ and $U(1)$ gauge scalars has the same structure as the coupling to the Standard Model gauge bosons. It is therefore useful to
define $\varphi_W^\pm = \frac{\varphi_{W_1} \pm i
  \varphi_{W_2}}{\sqrt{2}}$, $\varphi_Z = \cos \theta_W \varphi_{W_3}
+ \sin\theta_W \varphi_{B}$ and $\varphi_A= -\sin \theta_W
\varphi_{W_3}+ \cos\theta_W \varphi_{B}$. In this basis, the Higgs-induced mass Lagrangian is
\begin{equation}
  {\cal L}_{m_{EW}} =   -\frac{1}{8} g^2_{4D} v_{4D}^2 \left[
      \bar \varphi_W^+ \varphi_W^- + \bar\varphi_W^- \varphi_W^+ +
      \frac{\bar\varphi_Z \varphi_Z}{cos^2 \theta_W} \right]
\end{equation}
(we use the same convention that the presence
(absence) of bar means $m=-2$ ($m=2$) while the charge index is the
charge of the complete field).
As the structure of this mass term matches the one of the bulk mass, the spectrum can be described in terms of 4 complex fields: $\varphi_W^+$ (and its complex conjugate $\bar{\varphi}_W^-$), $\varphi_W^-$ (and $\bar{\varphi}_W^+$), $\varphi_Z$ ($\bar{\varphi}_Z$), and $\varphi_A$ ($\bar{\varphi}_A$), with masses:
\begin{equation}
m_{\varphi_W^+}^2 = m_{\varphi_W^-}^2 = \frac{6}{R^2} + m_W^2\,, \quad m_{\varphi_Z}^2 = \frac{6}{R^2} + m_Z^2\,, \quad m_{\varphi_A}^2 = \frac{6}{R^2}\,. 
\end{equation}

This mass term shows that the mass eigenstates after the Higgs
mechanism, at tree level, are in fact determined by the same change of
basis as in the Standard Model and that the mass contributions induced
are the same as the Standard Model corresponding gauge boson
masses. Hence at tree level, the photon scalar excitation is indeed
the dark matter candidate. 
Note that the mass splitting between the various states become very small in the limit of large radius, as they are proportional to the fixed value of the electroweak scale.
This picture may however be affected by large loop corrections, which are typically proportional to the bulk mass.

\subsubsection{Loop effects on the mass of $(2,\pm 2)$ are small}

In extra-dimensional models, the discussion of the dark matter
candidate implies computing the spectrum correction at one loop
because their effect largely dominates over electroweak effects
\cite{Cheng2002,Cacciapaglia2011}. We will however argue in this
section that the dominant loop mass splitting effects that could change the picture
displayed in the discussion of electroweak effects are absent from
level $(2,\pm 2)$. Using a 6D picture, we can write the part of the Lagrangian that determines the masses of Kaluza-Klein gauge scalars:

\begin{equation}
{\cal L}_{m_\varphi}=\frac{-1}{4} \left(f_\text{bulk}(\theta) F_{MN} F^{MN} + 2 f_\text{loc}\delta(\theta) F_{\theta\phi}F^{\theta\phi}\right)
\end{equation}

These terms have to be present and non-trivial to absorb all loop divergences appearing on the orbifold\footnote{The origin of these terms is best understood in the so called ``winding mode representation"  \cite{DaRold2004,Cheng2002} in which the extra-dimensional part of propagators is expressed in position space. This representation is useful on orbifolds because it allows for an simple expression of the propagator in terms of the propagators on the base space: it is a sum over all the propagators between the elements of the orbit of the end- and start--point: $G_\text{orb}(x,y)=\sum G(M_1 x, M_2 y)$. The localised term generation is easily illustrated by a tadpole diagram $\int dz G_\text{orb}(x,z)G_\text{orb}(z,z)G_\text{orb}(z,y)$, which has a divergent part which is $\sum\int dz G(x,M_i z) G(M_1 z,M_1 z) G(M_j z, y)$ for points in the bulk. Fixed points however, have additional divergent terms with the same degree of divergence because some $M_1\neq M_2$ verify $M_1 z_f = M_2 z_f$, which explains the need for localised operators to appear in the theory. 

The bulk, position-dependent term appears due to loops that explicitly break the symmetry of the base space and must therefore be related to some large distance effects mixed with UV divergence. This mixing is nicely illustrated by bubble diagrams. In general, a bubble is expressed as $\int dz_1 dz_2 G_\text{orb}(x,z_1)G^2_\text{orb}(z_1,z_2)G_\text{orb}(z_2,y)$ and its divergent part is contained on the part in which $z_1=z_2$. Cutting the legs, we are interested in $G^2_\text{orb}(z,z)$. Expanding over the winding modes, one finds that there are two kinds of divergent terms: $\propto G^2(z,z)$, which have the same structure as on the base space and are position independent, and $G(z,z)G(z,Mz)$, which have a subdominant divergent behaviour, have no equivalent in the bulk, and are position dependent because the distance between $z$ and $Mz$ depends on $z$ in general.} 
(see for example \cite{DaRold2004,Cheng2002,Cacciapaglia2012}). Because the localised terms are generated by more divergent loops than the bulk terms and can benefit from significant wave--function enhancements, they are expected to yield a dominant contribution. In practice, the bulk terms do cause some mass splitting but very mildly so, especially for states beyond the first KK tier \cite{Cacciapaglia2012}, which can be intuitively justified by their large-distance origin.

Because our dark matter candidate level has non-zero angular momentum $m$, its wave--function has to be zero on the fixed point, which means that there can be no effect of the dominant localised term on its mass. While this is not sufficient to claim that the mass eigenstates in the $(2,\pm 2)$ level are electroweak, we believe that there is a reasonable hint that the splitting is very moderate between states in the level. Due to the lack of a complete calculation of the one loop bulk terms, which is beyond the scope of the present simplified analysis, we will consider the very plausible hypothesis that the photon-like state is the lighter one. 
Although this choice of basis is grounded on quite weak arguments, a change of basis should not significantly alter the results of this study.  In the relic abundance calculation, both neutral states contribute if their mass splitting is smaller that the freeze-out temperature (which is typically in the GeV to tens of GeV range). Thus, under our assumption of small splitting, the choice of basis does not significantly affects the result.  For the LHC constraints we study in the next section, the mass splitting in the even tiers also becomes irrelevant because the large widths induced by the couplings to the localised fermions makes the two resonances merge into a single bump.

\section{A lower limit from di--lepton searches}
\label{dilepton}

A novel feature of models on spherical orbifold, compared to the flat case, is the presence of ``even'' states with mass equal or lower than the mass of the Dark Matter level. In the case under study, $S^2/S_4$, there is a level $(1,0)$ with mass $\sqrt{3}$--times smaller than the DM mass, and also containing gauge scalars. Those scalars do not couple directly to the localised fermions: even though their wave functions are not vanishing on the fixed point, gauge invariance only requires a coupling of the localised fermions to the polarisation of the gauge boson along the ordinary 4D. Thus, the $(1,0)$ scalars will dominantly be pair-produced via gauge interactions, which implies small rates. 
The level $(2,0)$, with the same mass as the Dark Matter one, is much more relevant for phenomenology: it contains spin-1 resonances of the gauge bosons together with Higgs ones. 
An important feature derives from our choice of localising the fermions: this implies that spin-1 resonances of the gauge bosons couple to the fermions with unsuppressed gauge couplings. This situation has to be contrasted to the case of bulk fermions, where the couplings are only loop induced. As a consequence, the resonances can be abundantly produced at the LHC via Drell-Yan. The most constraining channel is due to di--lepton resonances from the production of the $Z$ and photon resonances.

Both ATLAS \cite{ATLASCollaboration2014} and CMS \cite{CMSCollaboration2014} have reported the results of their Run I searches for such di--lepton resonances and provide both direct yields and limits on the parameter spaces of specific Standard Model extensions. While our model shares similarities with the Sequential Standard Model that both collaborations constrained, the differences due to the presence of an extra resonance (the KK photon) and rescaled couplings -- which impact widths significantly -- are too stark to reliably reinterpret their limits. Therefore, we have recast the ATLAS analysis -- whose lepton selection criteria are most compatible with the tools available to theorists -- to derive limits by comparing our simulations to their measurements.

We used FeynRules \cite{Alloul2013} to implement a Standard Model extension including
the two Kaluza-Klein vector resonances we wanted to investigate and
generated di--lepton events -- including the $Z$ boson contribution -- using
MadGraph5\_aMC@NLO \cite{Alwall2014} for parton-level simulations and PYTHIA \cite{Andersson2006} for
showering and hadronisation. The PYTHIA output was further processed to account for detector
effects using the MadAnalysis5 Delphes tune \cite{DeFavereau2014,Dumont2014}.

We used MadAnalysis5 \cite{Conte2012,Conte2014} at the reconstructed
level to recast the ATLAS analysis in the di--electron channel. This choice was motivated by the better performance reported by the ATLAS collaboration compared to the di--muon channel as well as the selection procedure differences between the two channels, which make MadAnalysis5 more suitable for a close reproduction of the di--electron channel selection criteria.
Table \ref{atlasvsus} compares our selection procedure with
that of the ATLAS analysis for reference, and shows that it is possible to closely match all the cuts that were applied on the data.
\begin{table}[!h]
\centering
  \begin{tabular}{|c|l|c|}
\hline Category & ATLAS & Recast ? \\ \hline
Electron trigger & 2 ECAL clusters&No\\
& consistent with E\&M showers,&\\
&$p_{T1}>\unit{35}{\giga\electronvolt}$,
$p_{T2}>\unit{25}{\giga\electronvolt}$&\\\hline
Two electrons & $\geq 2$ reconstructed electrons & Yes \\
& no charge requirement &   \\\hline
Electron rapidity &$|\eta| < 2.47$  & Yes \\
&$|\eta|<1.37$ or $|\eta|>1.52$&\\\hline
Electron &$E_{T_1} \geq \unit{40}{\giga\electronvolt}$ & Yes
 \\
transverse energy& $E_{T_2} \geq \unit{30}{\giga\electronvolt}$&\\\hline
Leading electron & $\sum E_{T} \leq 0.007 E_{T_1}
+\unit{5.0}{\giga\electronvolt}$& Yes \\
isolation & within $\Delta R \leq 0.2$ & \\\hline
Subleading& $\sum E_{T} \leq 0.022 E_{T_2}
+\unit{6.0}{\giga\electronvolt}$& Yes  \\
electron  isolation & within $\Delta R \leq 0.2$ &\\\hline
  \end{tabular}
  \caption{Cut flow of the di--lepton search: all criteria are easily reproducible using MadAnalysis5, except for the trigger.}
  \label{atlasvsus}
\end{table}
The selected events are then sorted into six di--lepton invariant mass
bins: \gev{110-200}, \gev{200-400}, \gev{400-800}, \gev{800-1200},
\gev{1200-3000}, \gev{3000-4500}.

\begin{figure}[!h]
  \centering
\includegraphics[width=0.8\textwidth]{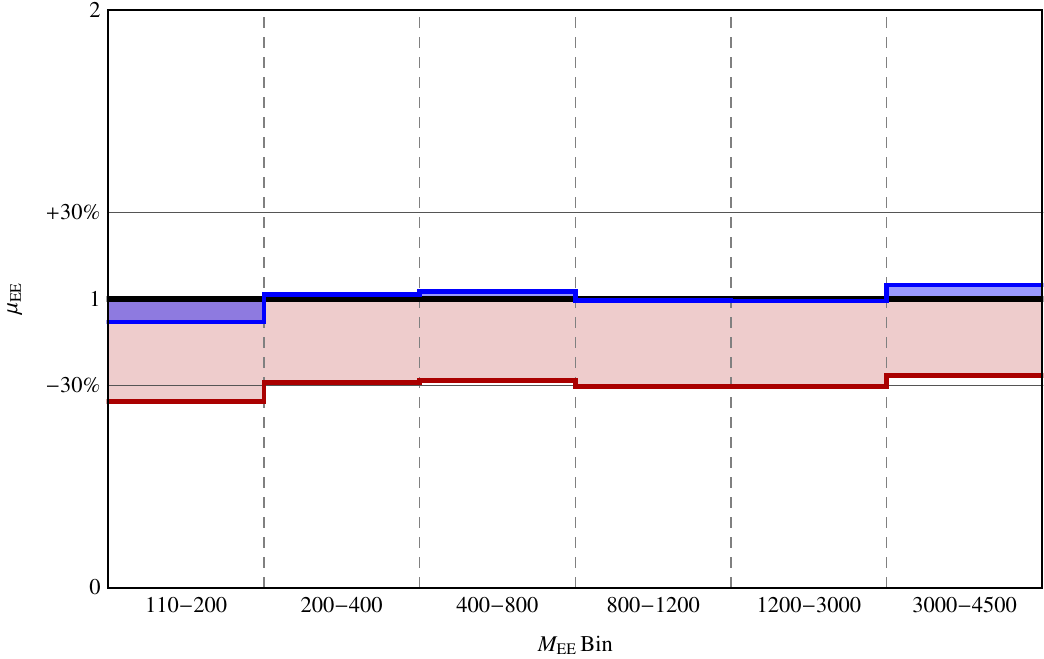}
\caption{Bin--per--bin ratios of Standard Model yields between our simulation and ATLAS's predictions, including a $K$-factor (red) and
normalised to ATLAS (blue)}
\label{picnorm}
\end{figure}

The quality of our implementation was controlled using a Standard Model-only sample generated with the same procedure as the
signal samples and the yields in each of the six bins were compared with the background estimates from ATLAS for the $Z$ boson background
channel. As shown in Figure \ref{picnorm} the results fall close to the $30\%$ accuracy advertised by the MadAnalysis5 recasting
guidelines \cite{Conte2014} once a 1.27 $K$-factor (determined using aMC@NLO) is included. We determined a scaling factor for our simulations 
by taking the average of the ratios between our prediction and that of ATLAS -- which is normalised to data. Apart from this overall normalisation, 
we can reproduce the ATLAS analysis Standard Model prediction to good accuracy.

\begin{figure}[!h]
\centering
\includegraphics[width=1\textwidth,angle=0]{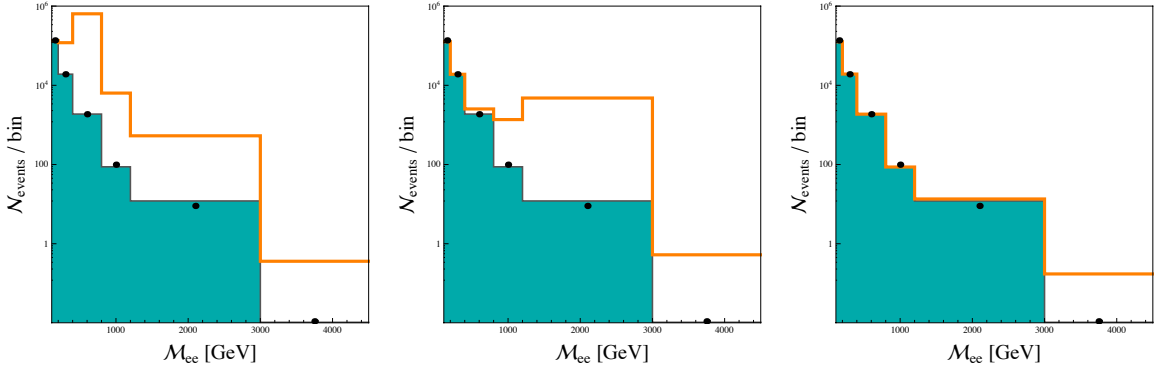}
\caption{Dielectron invariant mass spectra comparison between ATLAS data (black dots), the Standard Model prediction (blue) and the prediction of our model at different values of $1/R$ (\tev{250}, \tev{800} and \tev{1600}). As we will show, the first two values are experimentally excluded while the last is compatible with the ATLAS result.}
\label{thvsexp}
\end{figure}

Using this normalisation, we can now compare our Standard Model prediction to the ATLAS data and our prediction for signals at different compactification radii. As shown in Figure \ref{thvsexp}, the mass limit is expected to sit in the multi-TeV range, which we confirm using a statistical analysis.

We set a limit on the compactification radius by computing the confidence levels (CLs) in
each bin for our samples and we consider as excluded any sample where
at least one bin is more than $2\sigma$ away from the observed number
of events. The CLs are computed with the following
prescription:
\begin{itemize}
\item the signal is defined to be the yield obtained by processing our
  samples, which include the contribution from $Z$ bosons and
  interference effects.
\item the signal uncertainty is computed approximately. For the actual
  extra-resonances, the ATLAS analysis uses a bin-independent $4\%$
  systematic uncertainty, which strongly dominates the statistical
  uncertainty, which we always neglect. We need however to 
  include also the systematic uncertainty for the $Z$ boson contribution in
  our sample, which we do by adding in quadrature the uncertainty
  provided by ATLAS to the $4\%$ contribution which we compute from
  our simulations. This implies a double counting of some factors in
  the $Z$ background uncertainty, which is however small as the total
  uncertainty provided by ATLAS is significantly larger.
\item the background yield and uncertainty used for computing the CLs
  confidence levels are the contribution of all other channels
  computed by ATLAS with the uncertainties added in quadrature.
\end{itemize}

\begin{rst}
This analysis allowed us to set an exclusion limit on the radius of the sphere:
\begin{equation}
\frac{1}{R} > \unit{1.6}{\tera\electronvolt}\,, \qquad M_{l=2}>\tevm{3.9}\,,
\end{equation}
which translates to a lower bound on the mass of the DM candidate $M_{DM} \sim M_{l=2}$. 
\end{rst}

\section{An upper bound from relic density}
\label{dmrelic}
Let us now turn to the phenomenology of the dark matter candidate. We have established in Section \ref{dmlevel} that, to the
level of approximation we consider, this candidate is the complex scalar excitation of the photon in tiers $(2,\pm 2)$, which we
dub the lightest rotating particle (LRP). In this section we proceed to compute the expected relic density of our LRP in
the universe today assuming the Standard Model of cosmology. In this model, dark matter is produced in the thermal bath of the hot
primordial universe and leaves thermal equilibrium when the expansion rate of the universe dominates annihilation rates, causing a density
freeze-out.\\

As the symmetry protecting the DM candidate is a U(1), additional contribution to the relic density may come from a primordial asymmetry \cite{Nussinov:1985xr,Petraki:2013wwa,Zurek:2013wia}. We will not consider this possibility any further, because, as we will see, the thermal production is already excessive. It should be however noted that the asymmetric DM character may become relevant if the candidate never enters thermal equilibrium during the Universe's thermal history.

\subsection{Relic density calculations}

The calculation of the relic abundance of a stable particle species from thermal freeze-out is now a standard technique in dark matter
theory. The density $n$ of a dark matter particle well isolated in mass from other particles obeys the Boltzmann equation
\begin{equation}
  \label{eq:dmrelic1}
  \pd{n}{t} + 3Hn = - \avg{\sigma v} \left( n^2 - n_{eq} ^2\right)
\end{equation}
where $H$ is the expansion rate of the universe,
\begin{equation}
  \label{eq:dmrelic2}
  n_{eq}=g\left(\frac{mT}{2\pi} \right)^{\frac{3}{2}} e^{-m/T}
\end{equation}
is the thermal equilibrium density for a particle of mass $m$ with $g$
degrees of freedom and $\avg{\sigma v}$ is the thermally averaged
self-annihilation cross-section of dark matter pairs into Standard
Model particles. This equation can be used to derive at which
temperature $T_f$ dark matter ceases to be in thermal equilibrium and
essentially is no longer produced nor annihilated. The dark matter
density fraction is then expressed in terms of this temperature, the
quadratic speed expansion of the cross section $\avg{\sigma v} = a + b
\avg{v^2}$, the Planck mass $M_{pl}$ and the number of relativistic
degrees of freedom at freeze-out $g_*$:
\begin{equation}
  \label{eq:dmrelic3}
  \Omega_{DM} h^2 \approx \frac{1.04\times 10^9}{M_{pl}} \frac{m}{T_f
    \sqrt{g_*}} \frac{1}{a+3mb/T_f}\,.
\end{equation}

It has however been noted \cite{Griest1991} that this picture should be corrected in the
presence of other particles that have a mass very close to that of the
dark matter and eventually decay to the dark matter particle. In this
case, the near-degenerate states should all be taken into account in
the Boltzmann equation to
have a precise prediction of the dark matter density. The same authors
provide a simple way of treating all co--annihilation effects by noting
that the total density $N$ of all near-degenerate modes should
approximately be described by a Boltzmann as all the modes will
eventually decay to dark matter. The effective Boltzmann equation is
\begin{equation}
  \label{eq:dmrelic4}
  \pd{N}{t} + 3HN = - \avg{\sigma_{eff} v} \left( N^2-N_{eq}^2\right)\,,
\end{equation}
where the effective collision rate is a weighted sum of all
co--annihilation processes, accounting for Boltzmann factors:
\begin{equation}
  \label{eq:dmrelic5}
  \sigma_{eff}=\sum_i^K\sum_j^K \sigma_{ij}
  \frac{g_ig_j}{g_{eff}^2}(1+\Delta_i)^{3/2}(1+\Delta_j)^{3/2}
  e^{-x(\Delta_i + \Delta_j)}\,.
\end{equation}
In this definition, the sums go through all species entering in $N$ (the dark matter particle and all near-degenerate states),
$\sigma_{ij}$ is the cross-section for the process $ij\rightarrow SM$, $g_i$ is the number of degrees of freedom of state $i$,
$\Delta_i=\frac{m_i-m_{DM}}{m_{DM}}$ and 
\begin{equation}
  \label{eq:dmrelic6}
  \sum_i^K g_i(1+\Delta_i)^{3/2}e^{-x\Delta_i}\,.
\end{equation}
The effect of co--annihilations are in particular very important in many
Kaluza-Klein dark matter models \cite{Servant2003a,Arbey2012a} and as
we will see our model makes no exception, all the more as there is a
high degeneracy between the LRP and other modes in the same tier.

For our relic density predictions, we used FeynRules \cite{Alloul2013} to define models
from the interaction terms we derive in Appendix \ref{app:intlagrangian}. These FeynRules models
were used to generate CalcHEP models, which were then used in CalcHEP itself \cite{Belyaev} to
produce Mathematica \cite{WolframResearchInc} outputs for the relevant
cross-sections, using which we could go through the calculations
outlined in this section to obtain the present-day dark matter
density, using dark matter annihilations and co--annihilations.

This calculation can be used to put upper bounds on the mass of the
dark matter particle and thus, due to the competition with the
di--lepton searches, contribute to constrain the model. Indeed,
the relic density prediction typically grows with the mass of the dark
matter particle \cite{Kong2005},  and it has to be compared with the latest Planck
result: $\Omega h^2 = 0.1188 \pm 0.0010$ \cite{PlanckCollaboration2015}. This
measurement gives a preferred value for the inverse radius, at which
all the dark matter content in the universe can be consistently
explained by the model's LRP and any excess larger than two
standard deviations can be excluded. If the prediction is below the
measured value, the model cannot be excluded with certainty as dark
matter may have multiple components. However, larger values are ruled
out as it would lead to the overclosure of the Universe.

\subsection{Thermal relic density results}

In the following, we will consider two extreme cases: first, the case where only one state contributes to the DM relic abundance, and later we consider all the co-annihilation processes. The latter is a more realistic case, as the mass splittings we expect in this model are very small. We numerically checked that effects due to the splitting induced by the electroweak symmetry breaking are negligible, so that the results are shown in the most favourable case where such splittings are set to zero. The case without co-annihilation can be seen as a least favourable scenario.

\subsubsection{Annihilation process}

There is only one relevant channel for the annihilation of dark
matter: $\varphi_A \bar \varphi_A \rightarrow W^+ W^-$. Due to phase space
effects, the production of $W$ excitations is largely suppressed and can therefore be
neglected. Using the CalcHEP analytical expression for the matrix
element, we could compute the non-relativistic expansion of the
associated $\avg{\sigma v}$, which in turn allowed us to predict the dark matter density in our model, taking into account only
annihilation. 

This prediction is formulated as a function of the
radius of the extra-dimension, and $\Omega h^2$ increases with the mass scale of the
extra-dimensions as expected. The result is presented in Figure
\ref{fig:dmrelicann} where the Planck preferred value is represented
to show that the upper bound on the inverse radius inferred from this
calculation would be $\frac{1}{R} \leq \gevm{100}$ which in turns means
$ m_{l=2} \leq \gevm{250}$. This limit, however is too stringent as
co--annihilation effects with other gauge scalars change the prediction
for the relic density significantly.

\subsubsection{Co--annihilation processes}

\begin{figure}[!h]
\centering
\includegraphics[width=0.8\textwidth]{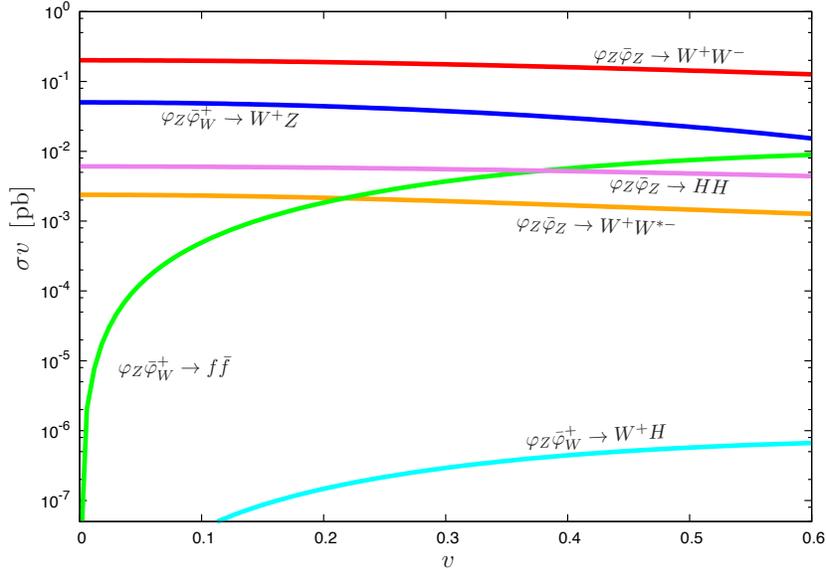}
\caption{Characteristic sample of annihilation channels}
\label{fig:coannsigv}
\end{figure}

\begin{figure}[!h]
  \centering
  \subfigure[$\varphi\bar\varphi\rightarrow V^\mu V^\mu$]{\includegraphics[width=0.2\textwidth]{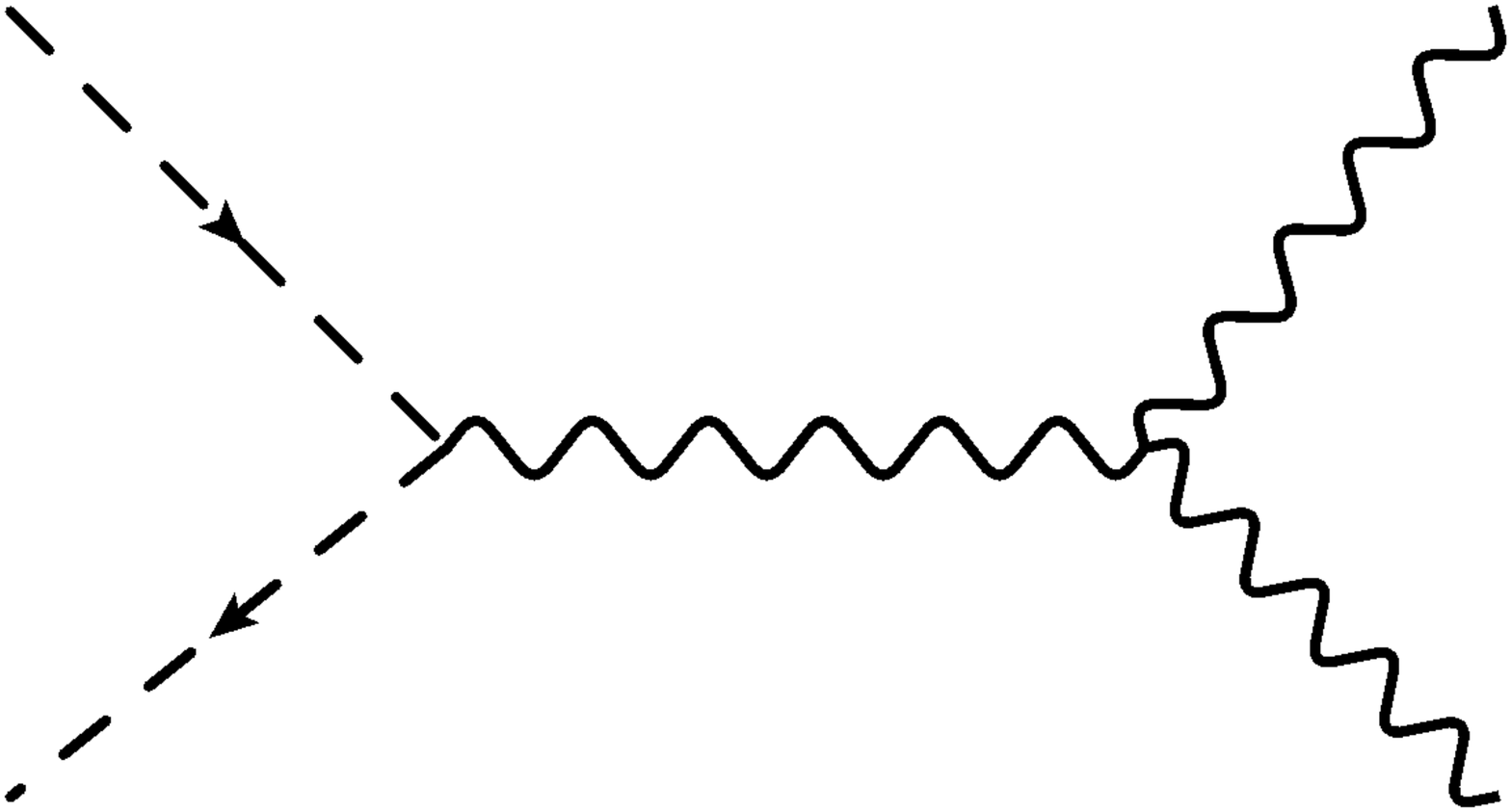} \label{fig:ddVV}}
  \hspace{1em}
  \subfigure[$\varphi\bar\varphi\rightarrow f\bar f$]{\includegraphics[width=0.2\textwidth]{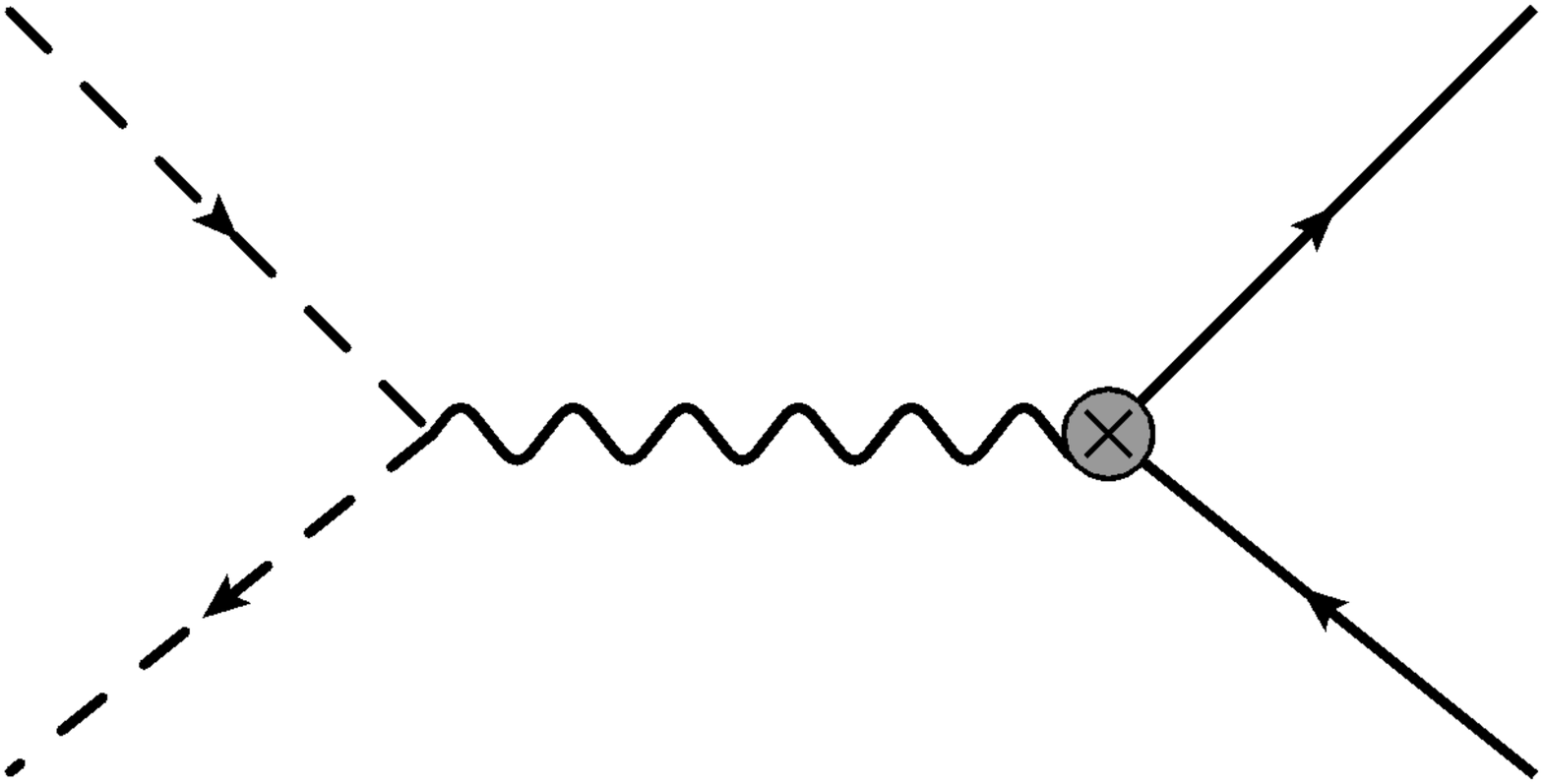}\label{fig:ddff}}
  \hspace{1em}
  \subfigure[$\varphi\bar\varphi\rightarrow V^\mu V^{*\mu} $]{\includegraphics[width=0.2\textwidth]{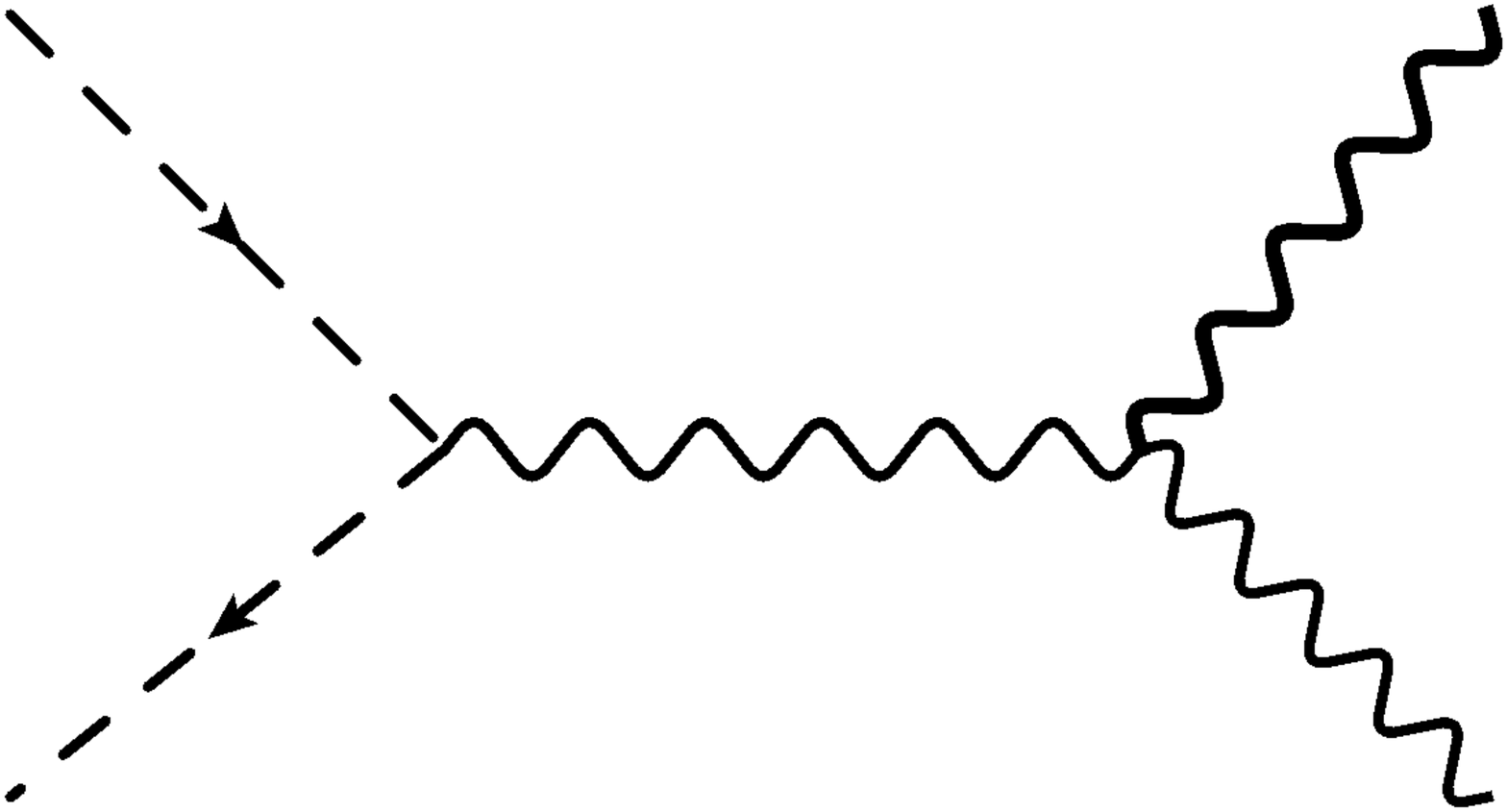}\label{fig:ddVVs}}
  \newline
  \subfigure[Vector Resonance]{\includegraphics[width=0.2\textwidth]{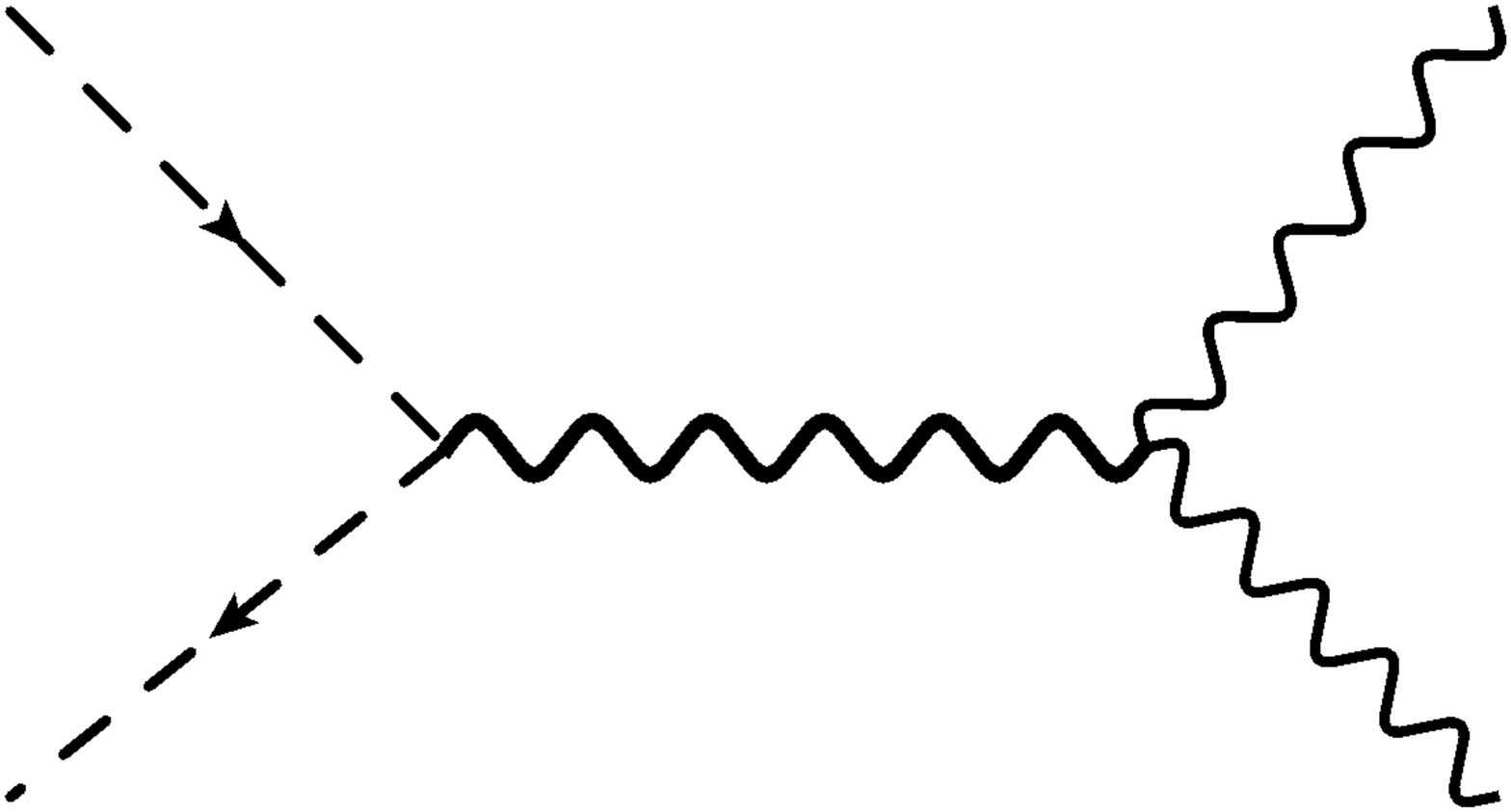}\label{fig:ddVV40V}}
  \hspace{1em}
  \subfigure[Higgs Resonance]{\includegraphics[width=0.2\textwidth]{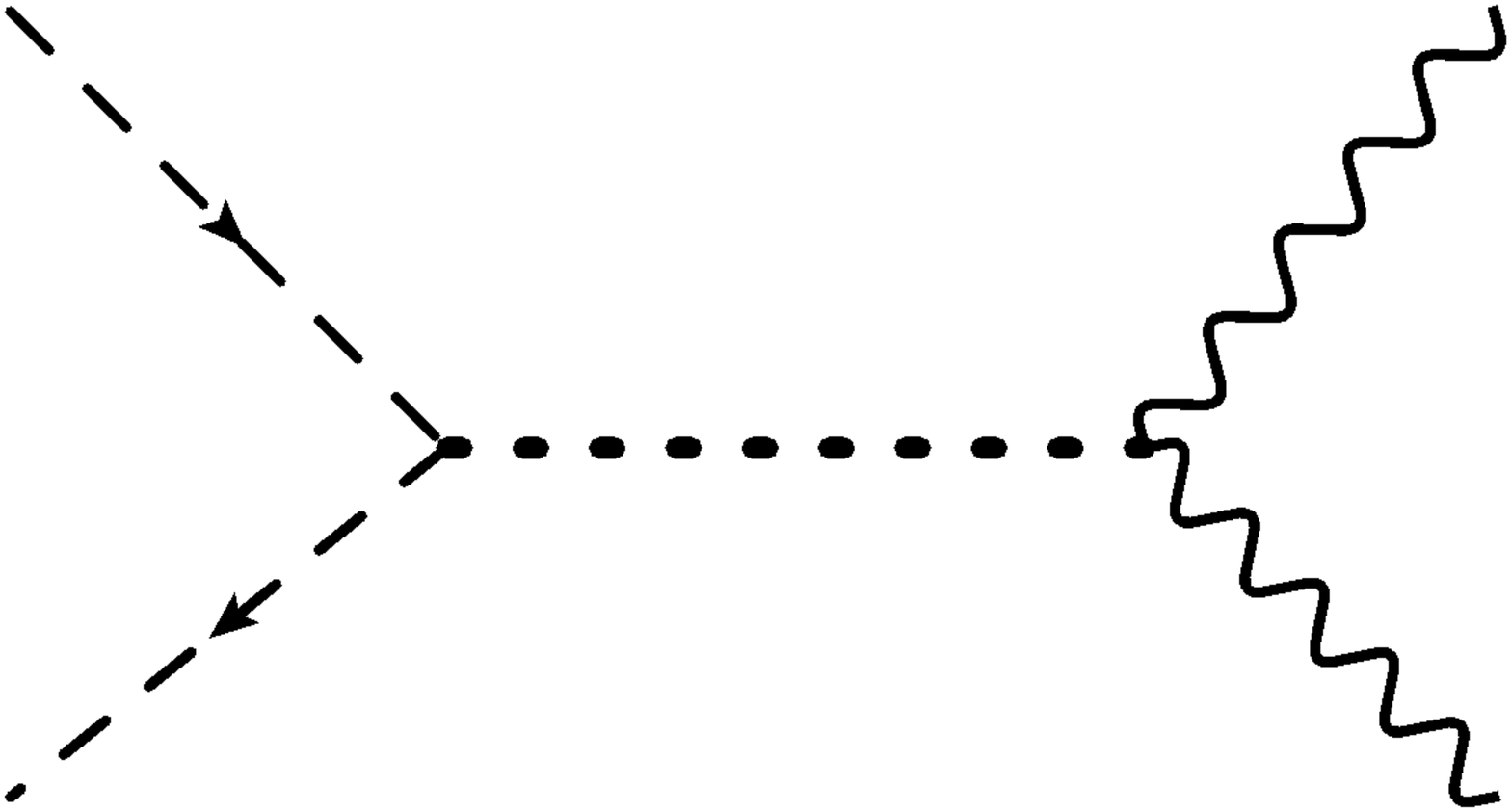}\label{fig:ddVV40H}}
  \caption{Examples of classes of processes. Thick lines represent Kaluza-Klein excitations and the fermion gauge coupling is marked to highlight its localised feature. 
eOur complete calculation also includes 4-point interactions for the relevant processes.
}
  \label{fig:dmgraphs}
\end{figure}

As we show in the Appendices \ref{app:intlagrangian} and \ref{app:channels}, there are several
channels available for co--annihilation of our dark matter candidate,
once the other electroweak gauge bosons are included in the
calculation. 
A list of all the processes we considered can be found in \ref{app:channels} and we show in Figure \ref{fig:coannsigv} a sample of thermal cross-sections computed using the dedicated CalcHEP module.
These processes can be sorted into several classes, as shown in Figure \ref{fig:dmgraphs}. As we will show below, the dominant contribution comes from co--annihilation to pairs of tier-0 vector bosons (Figure \ref{fig:ddVV}). Annihilation to fermions (Figure \ref{fig:ddff}) is a lot less important due to the absence of s-wave processes and the associated production of a vector boson and excited vector bosons (Figure \ref{fig:ddVVs}) is phase-space suppressed. 

It is worth mentioning that we considered the possible contribution of near-resonant s-channel diagrams involving $(4,0)$ excitations like in Figure \ref{fig:ddVV40V}, but that the resonance is still too far-off to have a significant impact on the thermal cross-sections. We should however note that a significant change can be introduced by fine-tuning the localised Higgs mass operator so that Higgs excitations hit precisely the resonance (Figure \ref{fig:ddVV40H}). A quick estimate using a subset of all processes showed that this changes significantly the thermal cross-section -- possibly providing a small funnel in parameter space where Dark matter is allowed to be very heavy. We however leave aside the study of this fine-tuned possibility in the present work. 

\begin{figure}[!h]
  \centering
\includegraphics[width=0.6\textwidth]{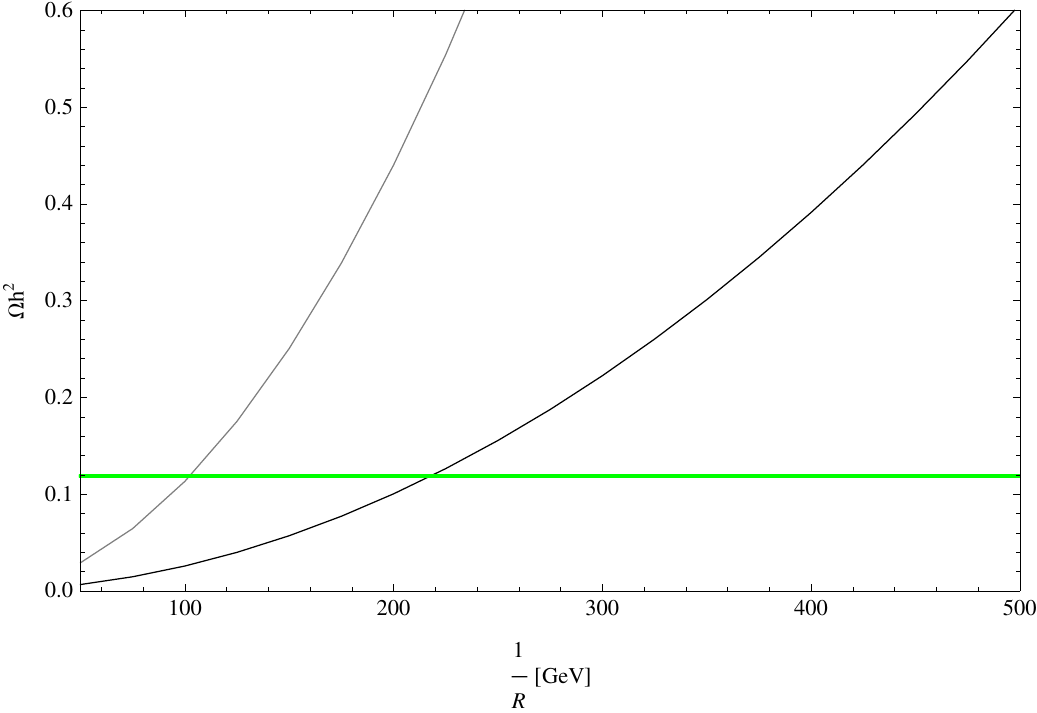}
  \caption{Dark matter relic density as a function of the sphere inverse
radius with (black) and without (grey) co--annihilations. The green band is centred around the
Planck preferred value and its width is given by the result's standard deviation}
\label{fig:dmrelicann}
\end{figure}

With the caveat of the possible Higgs resonance funnel, we could compute $\avg{\sigma v}_{eff}$ analytically and predict
for $\Omega_{DM}h^2$ as a function of $\frac{1}{R}$. The result of
this calculation is presented in Figure \ref{fig:dmrelicann}.
As expected, the upper value on $\frac{1}{R}$ is increased
significantly once co--annihilations are included. We now have the
constraint $\frac{1}{R}\leq \gevm{220}$, which translates to
$m_{l=2}\leq \gevm{540}$. While the tension is reduced with the
collider lower bound, there is no doubt that it is sufficient to rule
out the bulk of this model's parameter space, outside of a fine-tuned region.

\section{Other orbifolds}
Can we generalise the exclusion of this model to the other possible spaces? As we stated in Section \ref{orbifolds}, there are four families of non-polyhedric orbifolds that have fixed points and symmetries and, as we will see, we can use our result to draw the broad conclusion that none provides an experimentally viable framework for extending the Standard Model.
\begin{table}[!h]
\begin{center}
\begin{tabular}{|l|cc|cc|cccc|}
\hline
Orbifold & \multicolumn{2}{c|}{Dark Matter} & \multicolumn{2}{c|}{Z's} & \multicolumn{4}{c|}{wavefunction factors}\\
    &   $(l,m)$  &  spin  &  $(l,m)$  &  $m_{DM}/m_{Z'}$ & NP & SP & EME & WME \\\hline
$S_4$ & $(2,\pm2)$ & S & $(2,0)$ & $1$ & \multicolumn{2}{c}{$\sqrt{5}$} & - & - \\
$S_6$ & $(3,\pm3)$ & V & $(2,0)$ & $\sqrt{2}$ & \multicolumn{2}{c}{$\sqrt{5}$} & - & -  \\\hline
$C_2$ & $(2,\pm2)$ & V+S & $(1,0)$ & $\sqrt{3}$ & $-\sqrt{3}$ & $\sqrt{3}$ &-  & - \\
$C_4$ & $(4,\pm4)$ & V+S & $(1,0)$ & $\sqrt{10}$ & $-\sqrt{3}$ & $\sqrt{3}$ &-  & -\\\hline
$C_{2h}$ & $(2,\pm2)$ & V & $(2,0)$ & $1$ & \multicolumn{2}{c}{$\sqrt{5}$} & - & -  \\
$C_{4h}$ & $(4,\pm4)$ & V & $(2,0)$ & $\sqrt{10/3}$ & \multicolumn{2}{c}{$\sqrt{5}$} & - & -  \\\hline
$D_3$ & $(4,3)$ & V+S & $(2,0)$ & $\sqrt{10/3}$ & \multicolumn{2}{c}{$\sqrt{5}$} & $-\sqrt{5}/2$ & $-\sqrt{5}/2$  \\
$D_5$ & $(6,5)$ & V+S & $(2,0)$ & $\sqrt{7}$ & \multicolumn{2}{c}{$\sqrt{5}$} & $-\sqrt{5}/2$ & $-\sqrt{5}/2$  \\\hline
\end{tabular}
\end{center}
\caption{Properties of DM and lightest $Z'$ in the lowest $n$ spherical orbifolds. For DM, S stands for gauge scalars and V for gauge vectors and Higgs. The wavefunction factors refer to the fixed points as listed in Table~\ref{tab:orbprop}.}
\label{tab:DMprop}
\end{table}
The argument we put forward goes as follows:
\begin{itemize}
\item[-] In all cases, there exists an even level with mass lighter or equal to the one of the DM candidate and containing a spin-1 gauge resonance with unsuppressed couplings to the localised fermions~\footnote{Even gauge scalars are exempt from LHC constraints as they do not couple to the localised fermions at tree level.}. Thus, the constraint from the di-lepton search always applies and requires the DM state to appear in the multi-TeV range.
\item[-] Our analysis shows that the preferred mass range for the DM is well below the TeV, barring Higgs resonant channels. In order to reach the multi TeV scale, the annihilation cross section would need to be strongly enhanced: for $m_{DM} = 5$ TeV, one would need a cross section larger by roughly a factor of 80.
\item[-] The case we studied is the least favourable one, as the DM candidate is a gauge scalar. In other cases, it may be a vector, thus receiving additional contribution to the annihilation into a pair of fermions. There are also cases where the stable level contains both vector and scalar gauge states: in any case, the needed enhancement to escape the LHC bound is very unlikely to be obtained without the help of an s-channel resonant contribution.
\end{itemize}
While our argument is qualitative, it leads us to expect that all of the cases of natural DM on the sphere is in tension with data. A summary of the main features of the first two low--$n$ orbifold for each class can be found in Table~\ref{tab:DMprop}: it shows that in most cases the $Z'$ is lighter than the DM and has unsuppressed couplings. Thus, the LHC always implies a strong indirect constraint on the DM mass. Furthermore, the situation is always worsened for increasing $n$.
In the following, we will give more details for each class of orbifold.

\paragraph{Higher order $\boldsymbol{S_{2n}}$.}

\begin{figure}[!h]
  \centering
  \subfigure[$S_4$]{\includegraphics[width=0.3\textwidth]{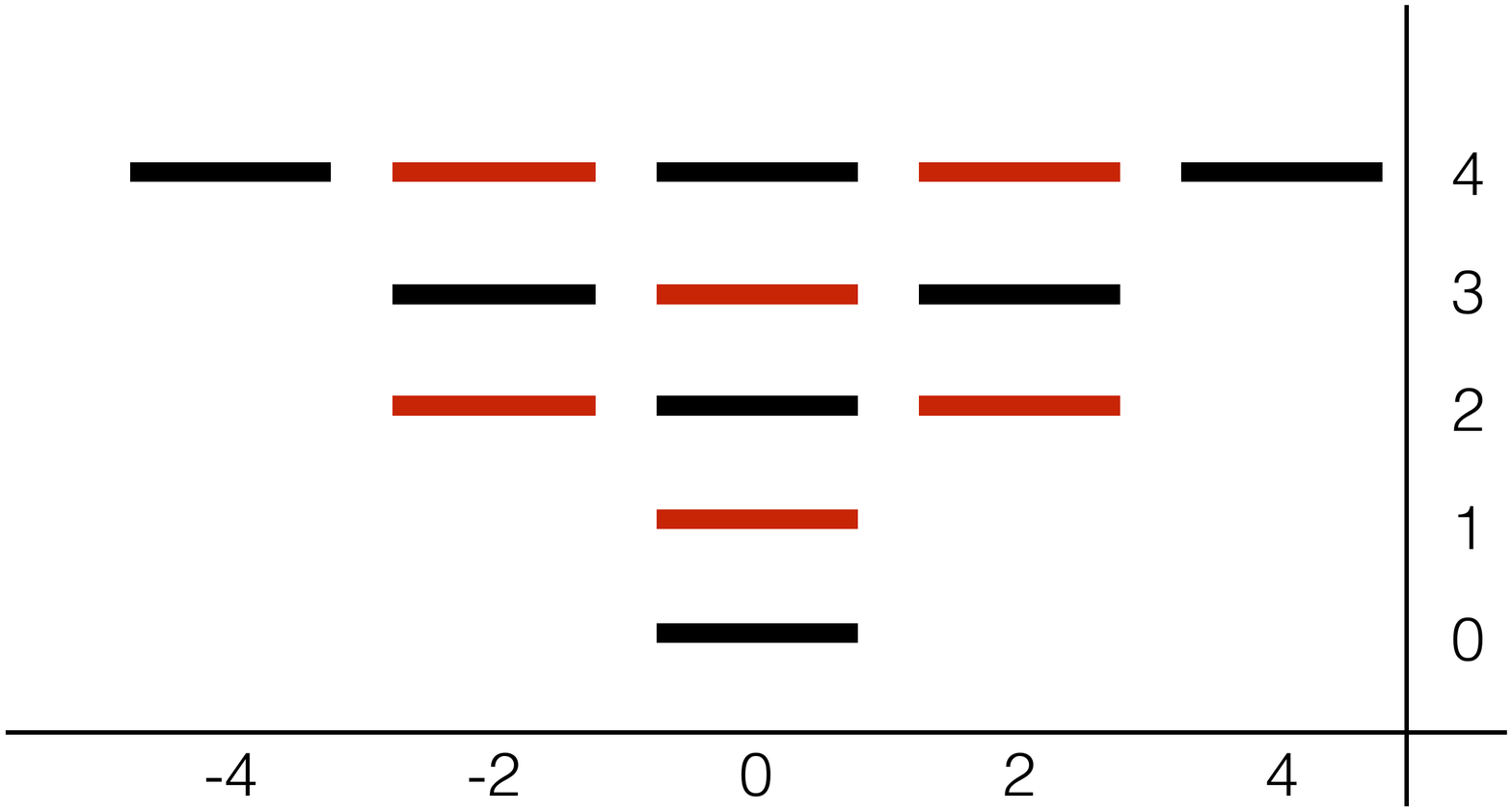}\label{fig:specs4}}
  \subfigure[$S_6$]{\includegraphics[width=0.3\textwidth]{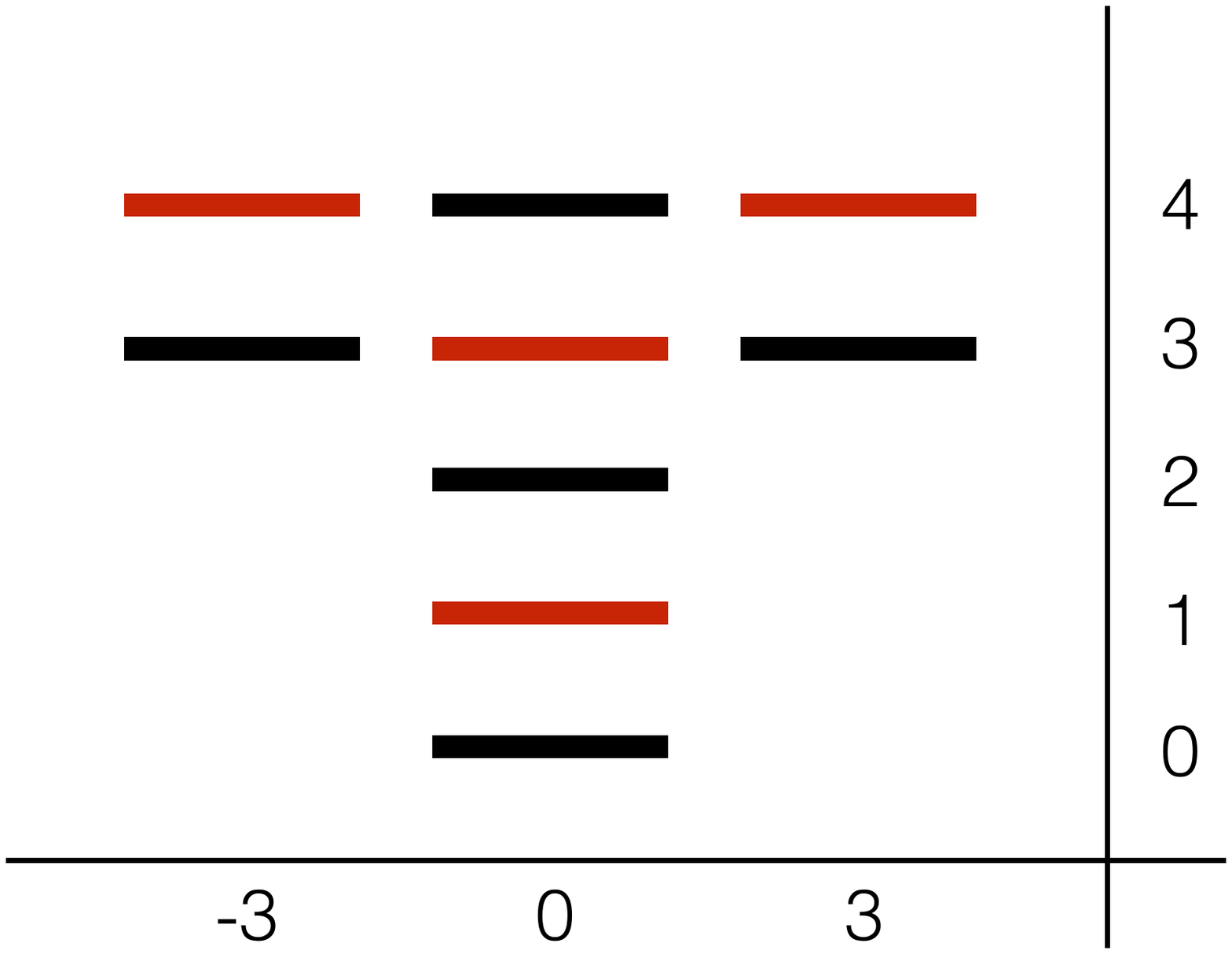}\label{fig:specs6}}
  \subfigure[$C_2$]{\includegraphics[width=0.3\textwidth]{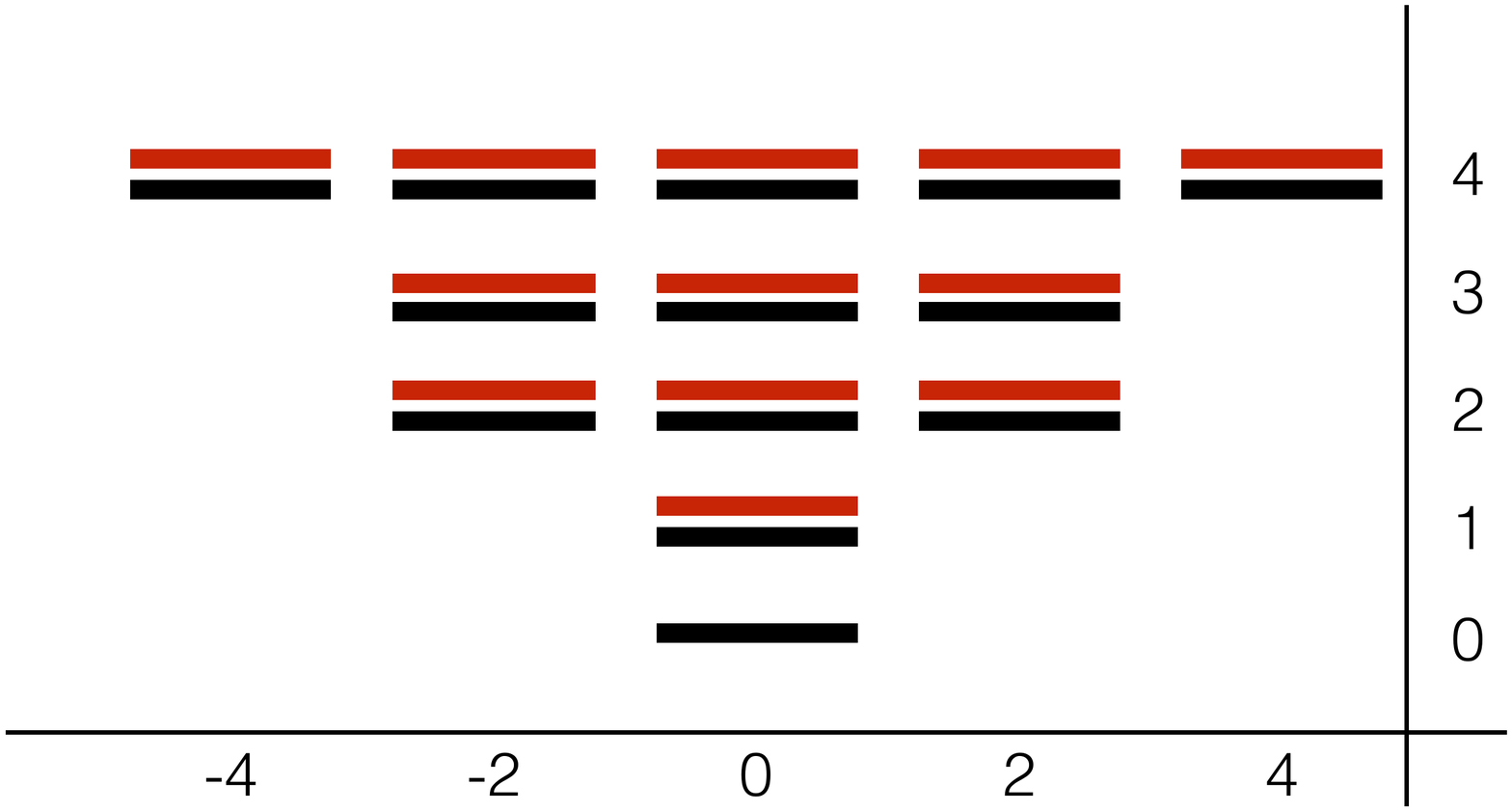}\label{fig:speccn}}
  \subfigure[$C_4$]{\includegraphics[width=0.3\textwidth]{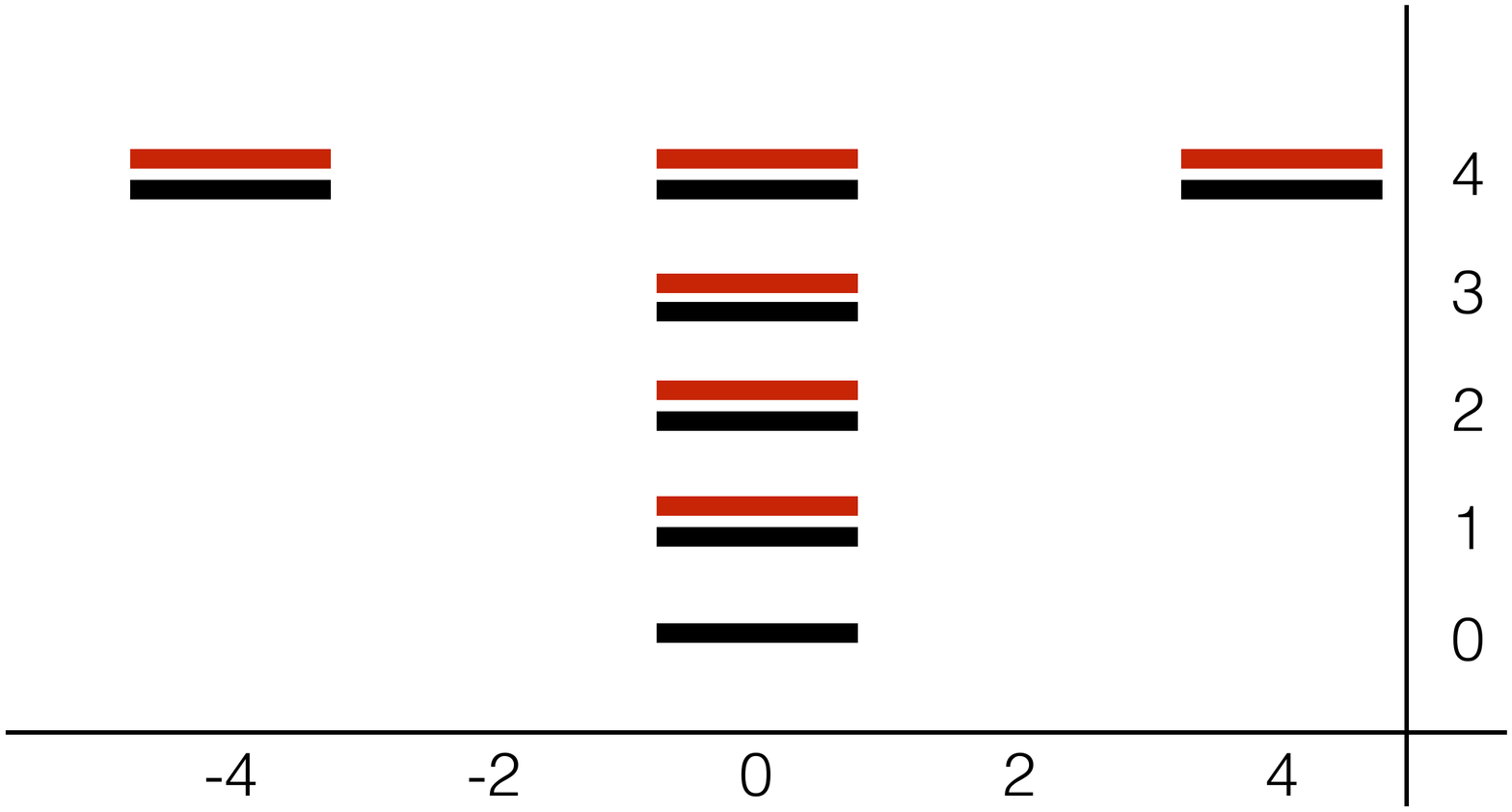}\label{fig:specc4}}
  \subfigure[$C_{2h}$]{\includegraphics[width=0.3\textwidth]{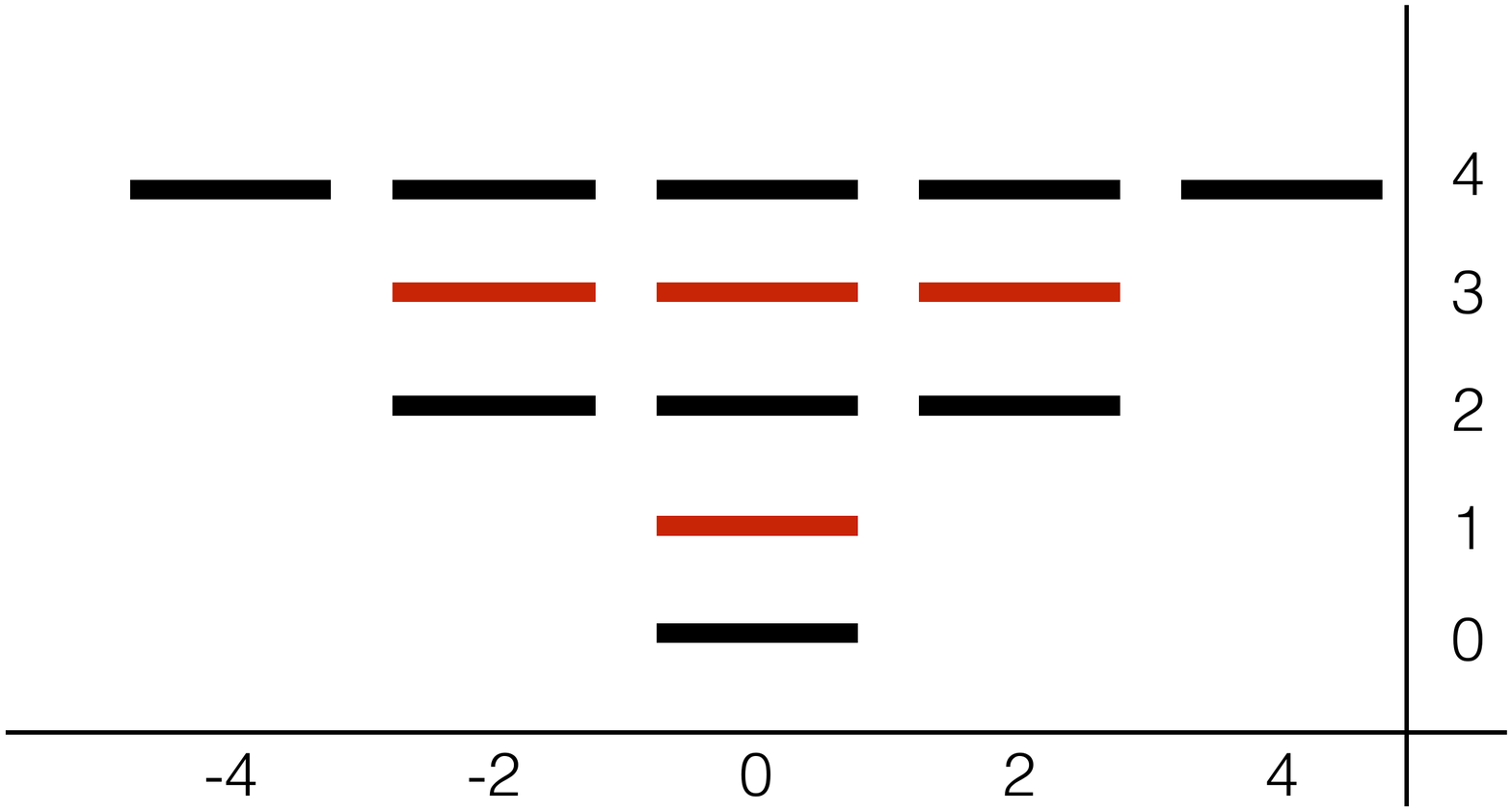}\label{fig:speccnh}}
  \subfigure[$C_{4h}$]{\includegraphics[width=0.3\textwidth]{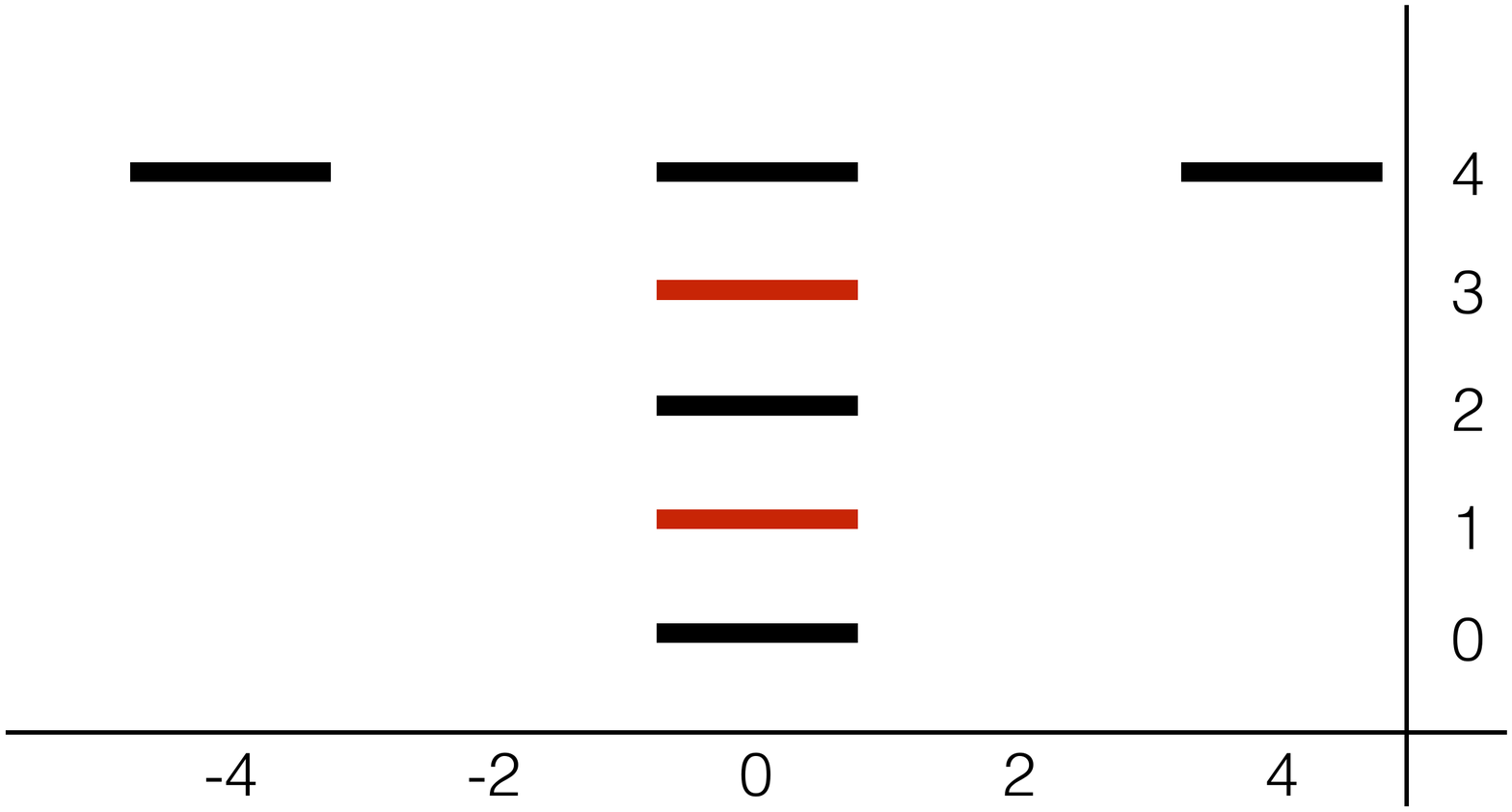}\label{fig:specc4h}}
  \subfigure[$D_{3}$]{\includegraphics[width=0.3\textwidth]{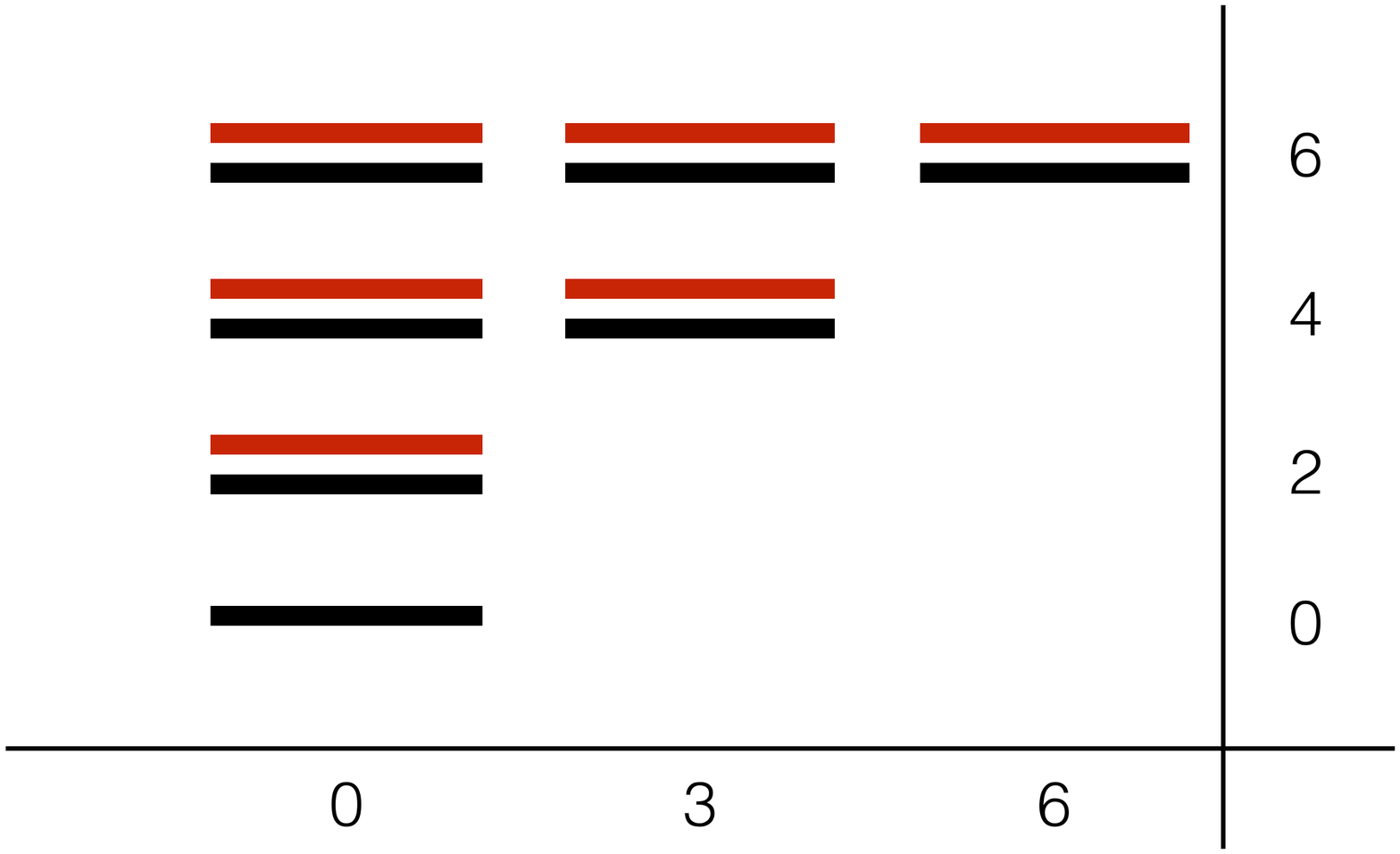}\label{fig:specdn}}
  \subfigure[$D_{5}$]{\includegraphics[width=0.3\textwidth]{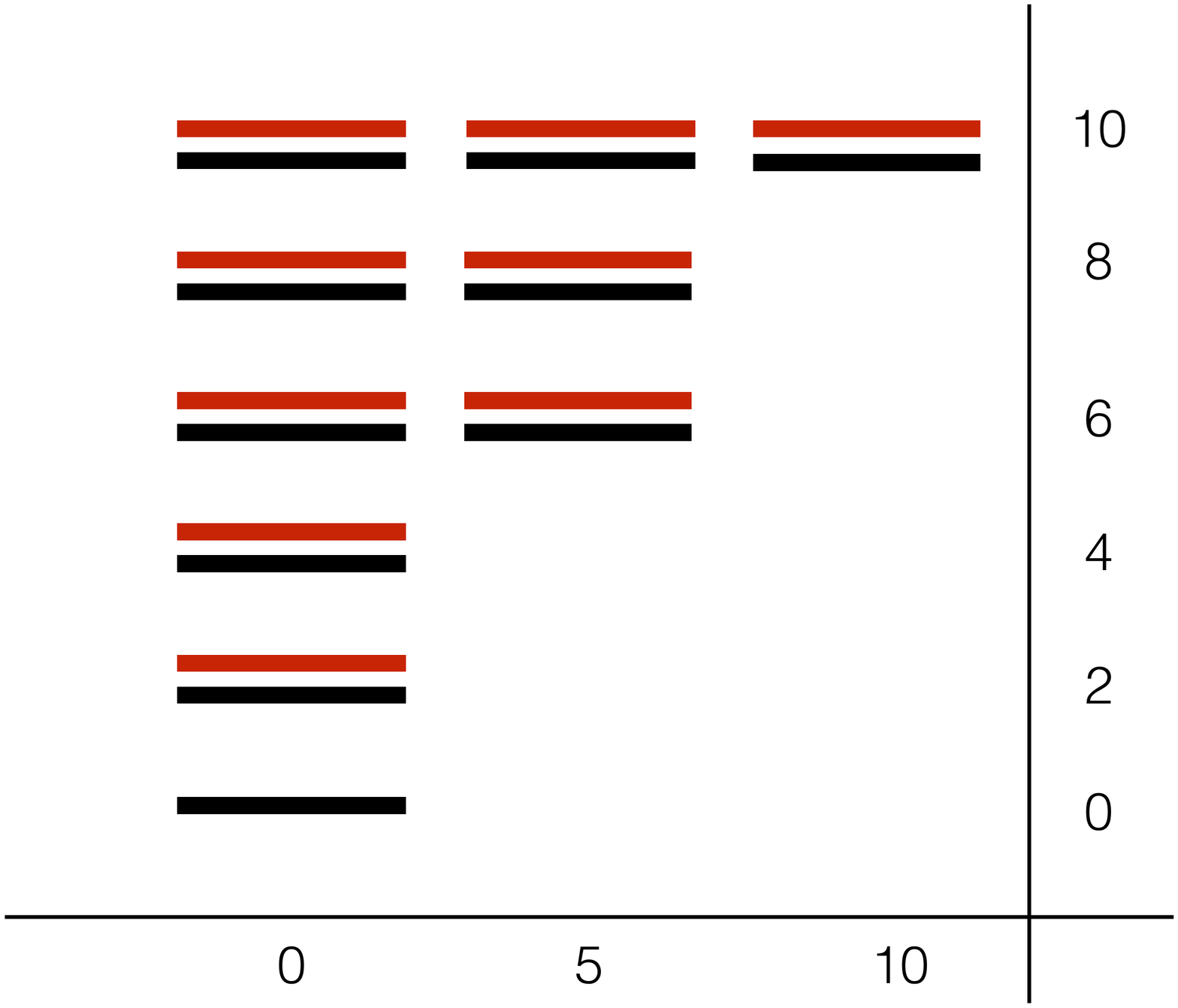}\label{fig:specd5}}
  \caption{Examples of spectra of gauge bosons on $S_n$,  $C_n$, $C_{nh}$, and $D_n$, with vectors levels displayed in black and scalars in red. The vertical axis spans values of $l$ while the horizontal axis spans $m$.}
\end{figure}

We gave a general-$n$ analysis of the spectrum of gauge bosons on $S^2/S_4$ in Section \ref{dmlevel}. For $n>2$, one can see that there will be 
several $m=0$ excitations lighter than the first stable level, which will appear for $l=n$, either in the vector or the scalar sector depending on the parity 
of $n$. In particular, the level $(2,0)$ always contains the lightest even $Z'$s. As an example, the spectrum of $S^2/S_6$ is shown in Fig.\ref{fig:specs6} ($S^2/S_4$ in Fig.\ref{fig:specs4} for comparison): the DM candidate is a spin-1 gauge boson in levels $(3,\pm3)$. 
The larger wavefunction factor and the fact that the DM candidate is heavier than the $Z'$ imply stronger bounds on the DM mass.

\paragraph{Cyclic groups $\boldsymbol{C_n}$.}
Spaces based on the groups $C_n$ have a $m$-conserving rotational symmetry, like $S_4$. 
As they are defined in terms of a rotation of $\phi$ by an angle $2\pi/n$, both vector and scalar gauge bosons have the same parities: thus each level, except the zero mode, contains both. Each mode $(l,m)$ will pick up a phase $e^{2 \pi i \frac{m}{n}}$, thus the modes that remain in the spectrum have $m/n$ integer.
This is illustrated in Figure~\ref{fig:speccn} for $C_2$ and Figure~\ref{fig:specc4} for $C_4$. Like for $S_{2n}$, the DM is contained in the lightest level with $m\neq 0$, i.e. $(n, \pm n)$, however such level contains both gauge scalars and vector (and the Higgs). This implies that the DM may have both spins, and that many states contribute to the co-annihilation processes and to the relic abundance. Nevertheless, the models always contain spin-1 gauge bosons with a mass lower than the DM one.
The lightest one is in $(1,0)$: even though its couplings to localised fermions are less enhanced than for $S_n$, c.f. Table~\ref{tab:DMprop}, the fact that it is substancially lighter than the DM candidate leads to a stronger bound on the DM mass than in $S_4$.
%
We reckon very unlikely that the enhanced co-annihilation would raise the preferred values of the DM mass to such values, even though a specific numerical study would be required to prove our intuition.

\paragraph{Horizontal cyclic groups $\boldsymbol{C_{nh}}$.}
Adding an extra horizontal mirror identification to $C_n$ to obtain $C_{nh}$ projects out more states out of the spectrum, as it now requires vectors to have even $(l+m)$ while scalars are odd (see Fig.~\ref{fig:speccnh} for $C_{2h}$ and Fig.~\ref{fig:specc4h} for $C_{4h}$). This projection preserves the symmetry of $C_n$, so $m$ is also conserved in this geometry. 
The DM level always contains gauge vectors and the Higgs resonances as the gauge scalars are projected out.
The lightest even spin-1 gauge boson is $(2,0)$, so that the LHC constraints are the same as for $S_4$: 
this implies $m_{DM} > 3.9$ TeV for $C_2$ and $m_{DM} > 7$ TeV for $C_4$, and increasing values for larger $n$.

\paragraph{Dihedral groups $\boldsymbol{D_n}$.}
The dihedral group is obtained by adding a $\pi$-rotation around an horizontal axis to $C_n$, which keeps even $l$ modes for both scalars and vectors and projects $m\neq 0$ states to the symmetric $Y_{lm}+Y_{l-m}$ configuration. There are two residual symmetries: a reflection around the horizontal axis and a reflexion about a vertical axis. The horizontal reflexion induces a phase $(-1)^{l+m}=(-1)^{m}$, and the vertical reflexion also induces a phase $(-1)^m$: thus, there is no odd state under the two parities for even $n$. 
The DM level is thus the level with odd $m$ and smallest possible $l$, i.e. $(n+1,n)$. The spectrum for $D_3$ is pictures in Fig.~\ref{fig:specdn}, while $D_5$ is represented in Fig.~\ref{fig:specd5}. Like the case of $C_n$, each tier, except the zero mode, contains both gauge scalars and vectors. 
The indirect LHC constraints are also determined by the gauge vectors in the level $(2,0)$, which couple directly to the localised fermions. However, in these models there are 3 independent singular points where the fermions may be localised: the poles (which are identified by the horizontal rotations), the west end of the meridian (WME at $\theta = \pi/2$ and $\phi = 0$), and the east end of the meridian (EME at $\theta = \pi/2$ and $\phi = \pi/n$). 
As it can be seen in Table~\ref{tab:DMprop}, the wavefunction enhancement factors on the equator fixed points are a factor os $1/2$ smaller than the $S_4$ case: this can ameliorate the indirect bounds from LHC, but not substantially as the DM level is always much heavier than the $Z'$, as it arises at $l\geq 4$.
We can thus conclude that the $D_n$ orbifolds are also disfavoured.

\section{Conclusion}
In this paper, we studied a class of sphere-based Extra-Dimension models with suitable geometries for featuring a dark matter candidate 
in the Kaluza-Klein spectrum. We selected the orbifolds that have a residual geometrical symmetry that prevents the decays of some resonances.
In order to obtain standard chiral fermions, instead of the Randjbar-Daemi--Salam--Strathdee mechanism used in previous works, we employ brane-localised fermions on orbifold fixed points. We provided a survey of all spherical symmetry 
groups, which provided a list of locally-spherical candidates for a Kaluza-Klein dark matter model with localised fermions. 
We constructed models where all fermions, and the gluons, are localised on a fixed point, while the electroweak gauge bosons and the Brout-Englert-Higgs field are allowed to propagate in the bulk. In general, the radius of the sphere will be bounded from above from overclosure of the Universe, while a lower pound is obtained by direct search at the LHC of even spin-1 resonances, in particular neutral $Z'$s decaying into a pair of leptons.
Focusing on a specific simple 
model based on $S^2/S_4$, we studied in detail both the LHC bounds and the relic abundance constraints. We found that the two constraints completely rule out the parameter space of the model. In our study we did not explicitly include the impact of localised operators on the fixed points of the orbifold. We argued that due to the peculiar properties of this model, where the dark matter level is insensitive to such localised operators, the only operator which could help release the tension between the dark matter and collider bounds is a localised Higgs mass term. Such an operator allows to fine-tune the mass of the Higgs excitations and hit a large coannihilation resonance, making KK dark matter possible in the\tev{} region. 

Besides this ``Higgs funnel'' region, the tension between the two constraints is such that we can extend the conclusion to all non-polyhedral orbifold with a Dark Matter candidate. After the Run-I of the LHC, therefore, models of Dark Matter with localised fermions, based on spherical orbifold, are strongly disfavoured.

\section*{Acknowledgements}
AD is partially supported by the ``Institut Universitaire de France''.  We also acknowledge partial support from 
the D\'{e}fiInphyNiTi - projet structurant TLF; the Labex-LIO (Lyon Institute of Origins) under grant ANR-10-LABX-66 and FRAMA 
(FR3127, F\'ed\'eration de Recherche ``Andr\'e Marie Amp\`ere").

\section*{Appendices}
\appendix
\section{Spherical orbifold groups}
\label{app:orbdetails}
In this section we provide a description of the non-polyhedral discrete subroups of $O(3)$ based on explicit coordinates for more clarity. We will use the spherical coordinates $(\theta,\phi)$ where $\theta$ is the colatitude and $\phi$ is the longitude. For each group, we will describe the action of the generators on these coordinate:

{\centering
\begin{tabular}{c c c}
\( \displaystyle \boldsymbol{C_n}:  \begin{matrix} (\theta,\phi) & \rightarrow & (\theta,\phi+\frac{2\pi}{n}) \end{matrix} \)  & 
\( \displaystyle \boldsymbol{C_{nh}}:  \begin{matrix} (\theta,\phi) & \rightarrow & (\theta,\phi+\frac{2\pi}{n}) \\ 
 (\theta,\phi) & \rightarrow & (\pi-\theta,\phi)
\end{matrix} \) &
\( \displaystyle \boldsymbol{C_{nv}}:  \begin{matrix} (\theta,\phi) & \rightarrow & (\theta,\phi+\frac{2\pi}{n}) \\
 (\theta,\phi) & \rightarrow & (\theta,-\phi) 
\end{matrix} \) \\
\phantom{a}&\phantom{a}&\phantom{a}\\
\( \displaystyle \boldsymbol{D_n}:  \begin{matrix} (\theta,\phi) & \rightarrow & (\theta,\phi+\frac{2\pi}{n})\\
 (\theta,\phi) & \rightarrow & (\pi-\theta,-\phi)
 \end{matrix} \)  & 
\( \displaystyle \boldsymbol{D_{nh}}:  \begin{matrix} (\theta,\phi) & \rightarrow & (\theta,\phi+\frac{2\pi}{n})\\
 (\theta,\phi) & \rightarrow & (\pi-\theta,-\phi)\\
 (\theta,\phi) & \rightarrow & (\pi-\theta,\phi)
 \end{matrix} \)  & 
\( \displaystyle \boldsymbol{D_{nd}}:  \begin{matrix} (\theta,\phi) & \rightarrow & (\theta,\phi+\frac{2\pi}{n})\\
 (\theta,\phi) & \rightarrow & (\pi-\theta,-\phi)\\
 (\theta,\phi) & \rightarrow & (\theta,\frac{2\pi}{n}-\phi)
 \end{matrix} \) \\
\phantom{a}&\phantom{a}&\phantom{a}\\
  &\( \displaystyle \boldsymbol{S_{2n}}:  \begin{matrix} (\theta,\phi) & \rightarrow & (\pi-\theta,\phi+\pi\left(1+\frac{1}{n}\right)) \end{matrix} \)  & 
\end{tabular}
}

The polyhedral subgroups $T$ and $O$ do not have a preferred axis and therefore have transformations which are quite complex when expressed in spherical coordinates. Their description is much easier by making their relation to polyhedra explicit

\begin{itemize}
\item $\boldsymbol{T}$ has the rotational symmetries of a tetrahedron: 4 axes going through a vertex and its opposing face and 3 axes going through the middle of two opposing edges.
\item $\boldsymbol{O}$ has the rotational symmetries of a cube: 3 axes going through the faces, 4 axes going through two opposing edges, 2 axes going through the diagonals.
\end{itemize}
\section{Lagrangian terms used for (co)annihilations}
\label{app:intlagrangian}
In this Appendix, we provide the expressions for the couplings we used in Section \ref{dmrelic} to derive a prediction for dark matter relic density in our $S^2/S_4$ model. We provide results in the manifestly symmetric basis without Kaluza-Klein expansion and the induced term for the physical eigenstates relevant to our calculations.

\subsection{Trilinear gauge couplings}

A relevant term for dark matter annihilation is the trilinear term coupling two gauge scalars to one gauge vector:

\begin{equation}
  \label{eq:1}
 {\cal L}_{SSV}=\frac{g}{2}\epsilon_{abc} A_\mu^a \left(
  \eth\Phi^b \partial_\mu \bar\eth\Phi^c + \bar \eth
  \Phi^b \partial\eth\Phi^c \right).
\end{equation}

From this expression we can derive the trilinear coupling of level
$(2,\pm 2)$ to the Standard Model gauge bosons:

\begin{equation}
  \label{eq:6}
  \begin{array}{rl}
    {\cal L}^{\sc{4D}}_{\sc{SSV}_1}= \left. g_{4D}\frac{1}{2i} \right[&
 \left(\partial_\mu\bar\varphi^+_W \varphi^-_W
      - \partial_\mu\bar\varphi^-_W \varphi^+_W\right) W^{3\mu}_{\sc{SM}}
\\
+&  \left(\partial_\mu\varphi^+_W \bar\varphi^-_W
      - \partial_\mu\varphi^-_W \bar\varphi^+_W\right)
    W^{3\mu}_{\sc{SM}}\\
-&\left( W^{+\mu}_{\sc{SM}} \varphi_W^- - W^{-\mu}_{\sc{SM}} \varphi_W^+
\right)\partial_\mu \bar\varphi_{W^3}\\
-&\left( W^{+\mu}_{\sc{SM}}\bar \varphi_W^- - W^{-\mu}_{\sc{SM}} \bar\varphi_W^+
\right)\partial_\mu \varphi_{W^3}\\
+&\left( W^{+\mu}_{\sc{SM}} \partial_\mu \varphi_W^- - W^{-\mu}_{\sc{SM}} \partial_\mu \varphi_W^+
\right)\bar\varphi_{W^3}\\
+&\left.\left( W^{+\mu}_{\sc{SM}}\partial_\mu \bar \varphi_W^- - W^{-\mu}_{\sc{SM}} \partial_\mu \bar\varphi_W^+
\right)\varphi_{W^3}\right]\\
  \end{array}
\end{equation}

and to the $(2,0)$ vector excitations:

\begin{equation}
  \label{eq:6}
  \begin{array}{rl}
    {\cal L}^{\sc{4D}}_{\sc{SSV}_2}= \left. g_{4D}\frac{\sqrt{5}}{14i} \right[&
 \left(\partial_\mu\bar\varphi^+_W \varphi^-_W
      - \partial_\mu\bar\varphi^-_W \varphi^+_W\right) W^{3\mu}_{(2,0)}
\\
+&  \left(\partial_\mu\varphi^+_W \bar\varphi^-_W
      - \partial_\mu\varphi^-_W \bar\varphi^+_W\right)
    W^{3\mu}_{(2,0)}\\
-&\left( W^{+\mu}_{(2,0)} \varphi_W^- - W^{-\mu}_{(2,0)} \varphi_W^+
\right)\partial_\mu \bar\varphi_{W^3}\\
-&\left( W^{+\mu}_{(2,0)}\bar \varphi_W^- - W^{-\mu}_{(2,0)} \bar\varphi_W^+
\right)\partial_\mu \varphi_{W^3}\\
+&\left( W^{+\mu}_{(2,0)} \partial_\mu \varphi_W^- - W^{-\mu}_{(2,0)} \partial_\mu \varphi_W^+
\right)\bar\varphi_{W^3}\\
+&\left.\left( W^{+\mu}_{(2,0)}\partial_\mu \bar \varphi_W^- - W^{-\mu}_{(2,0)} \partial_\mu \bar\varphi_W^+
\right)\varphi_{W^3}\right]\\
  \end{array}
\end{equation}

\subsection{Quadrilinear gauge couplings}

Quadrilinear gauge couplings also contribute to coannihilation processes. The unexpanded symmetric term is expressed below:

\begin{equation}
  \label{eq:3}
  {\cal L}_{SSVV} = \frac{1}{2}g^2 \left[ A_\mu^a A_a^\mu \eth \Phi^b \bar \eth \Phi_b
  - A^a_\mu \eth\Phi_a A_b^\mu \bar\eth \Phi^b  \right],
\end{equation}

which we can break down into different contributions.

\paragraph{Neutral scalars coupling to charged vectors}

The part ${\cal L}_{V_sV_sWW}$ which couples neutral gauge scalars and SM $W$ bosons are all
included in the first term of $\int{\cal L}_{SSVV}$:

\begin{equation}
  \label{eq:2}
    {\cal L}_{ V_sV_sWW}= g^2_{4D} W^+_{SM\mu } W_{-\mu}^{SM}
\bar\varphi_{A^3} \varphi_{A^3}
\end{equation}

\paragraph{Neutral and charged scalars coupling to
  neutral and charged vectors}
This mixed coupling ${\cal L}_{V_sW_sVW}$ arises from the second term in $\int{\cal L}_{SSVV}$:

\begin{equation}
  \label{eq:4}
      {\cal L}_{V_sW_sVW}=- \frac{1}{2}g_{4D} A^3_\mu \left[ \varphi^3 \left(W^{+\mu}\bar\varphi_W^-+W^{-\mu}\bar\varphi_W^+\right) + \bar\varphi^3 \left(W^{+\mu}\varphi_W^-+W^{-\mu}\varphi_W^+\right) \right]
\end{equation}

\paragraph{Charged scalars annihilation to charged vectors}
This coupling, ${\cal L}_{W_sW_sWW}$ requires both terms from $\int{\cal L}_{SSVV}$:

\begin{equation}
  \label{eq:5}
  \begin{array}{rl}
      {\cal L}_{W_sW_sWW}&= \frac{1}{2} g^2_{4D}
\left[ W_+^{SM\mu} W_{-\mu}^{SM} \left( \varphi_W^+ \bar\varphi_W^- +
    \varphi_W^- \bar\varphi_W^+ \right) - W^{SM}_{+\mu} W^{SM\mu}_{+}
  \bar\varphi_W^- \varphi_W^-\right. \\
&\left.\phantom{aaaaaaaaa aaaaaaaaa aaaaaaaaa aaaaaaa}- W^{SM}_{-\mu} W^{SM\mu}_{-}
  \bar\varphi_W^+ \varphi_W^+\right]
  \end{array}
\end{equation}

\paragraph{Charged scalars annihilation to neutral vectors}
This coupling, ${\cal L}_{W_sW_sVV}$ requires only the first term from
${\cal L}_{SSVV}$:
\begin{equation}
  \label{eq:7}
     {\cal L}_{W_sW_sVV} =\frac{1}{2}g^2_{4D} A^3_{SM\mu} A^{SM\mu_3} \left[\bar\varphi_W^+
  \varphi_W^- + \bar \varphi_W^- \varphi_W^+ \right]
\end{equation}

\subsection{Scalar gauge boson couplings to the Higgs boson}

In this section, we present the expression for the Higgs field couplings to the gauge scalars, as these provide a channel for dark matter coannihilation. The 6D gauge invariant expression for the term generating trilinear and quadratic couplings is 

\begin{equation}
  {\cal L}_{SSHH} = - \frac{1}{8} H^\dagger H \bar\eth \phi^a \eth \phi
\end{equation}

\paragraph{Quadratic coupling for scalar annihilation to Higgs boson pairs}

This coupling is obtained from $\int{\cal L}_{SSHH}$ by expanding in KK modes and projecting to physical states, which gives a 4D lagrangian term

\begin{equation}
  \label{eq:9}
      {\cal L}_{SSHH}^{4D} = -\frac{1}{8} g^2_{4D} hh \left[
      \bar \varphi_W^+ \varphi_W^- + \bar\varphi_W^- \varphi_W^+ +
      \frac{\bar\varphi_Z \varphi_Z}{\cos^2 \theta_W} \right]
\end{equation}

\paragraph{Trilinear coupling for scalar annihilation to a single Higgs boson}

We can easily generate the trilinear coupling of the dark matter level to Higgs bosons by replacing the Higgs field by its vacuum expectation value:

\begin{equation}
  \label{eq:8}
      {\cal L}_{SSH}^{4D} = -\frac{1}{4}g_{4D}^2 v h \left[ \bar \varphi_W^+ \varphi_W^- + \bar\varphi_W^- \varphi_W^+ +
      \frac{\bar\varphi_Z \varphi_Z}{cos^2 \theta_W}  \right] \\
\end{equation}
\newpage
\section{Annihilation and coannihilation channels}
\label{app:channels}
\renewcommand*{\arraystretch}{1.5}

\subsection{Vector Final States}
\begin{table}[!h]
\centering
\begin{tabular}{c|c|c||c|c|c}
Initial state & Final state & $\langle \sigma v \rangle$ & Initial
state & Final state & $\langle \sigma v \rangle$\\\hline
$\varphi_A\ \bar \varphi_A$& $W^+_\mu W^-_\mu $& $\sigv{1}$ &
$\varphi_A\ \bar \varphi_Z$& $ W^+_\mu W^-_\mu $ & $\sigv{2}$ \\
$\varphi_Z\ \bar \varphi_A$& $W^+_\mu W^-_\mu $&$\sigv{2}$ &
$\varphi_Z\ \bar \varphi_Z$& $ W^+_\mu W^-_\mu $ & $\sigv{3}$\\

$\varphi_A\ \bar \varphi_W^+$& $W^+_\mu A_\mu $&$\sigv{4}$ &
$\varphi_A\ \bar \varphi_W^-$& $ W^-_\mu A_\mu $ &$\sigv{4}$ \\
$ \varphi_W^+ \bar \varphi_A\ $& $W^+_\mu A_\mu $& $\sigv{4}$ &
$ \varphi_W^-\bar \varphi_A\ $& $ W^-_\mu A_\mu $ & $\sigv{4}$ \\

$\varphi_A\ \bar \varphi_W^+$& $W^+_\mu Z_\mu $&$\sigv{5}$ &
$\varphi_A\ \bar \varphi_W^-$& $ W^-_\mu Z_\mu $ & $\sigv{5}$ \\
$ \varphi_W^+ \bar \varphi_A\ $& $W^+_\mu Z_\mu $&$\sigv{5}$&
$ \varphi_W^-\bar \varphi_A\ $& $ W^-_\mu Z_\mu $ &$\sigv{5}$ \\

$\varphi_Z\ \bar \varphi_W^+$& $W^+_\mu Z_\mu $& $\sigv{6}$&
$\varphi_Z\ \bar \varphi_W^-$& $ W^-_\mu Z_\mu $ & $\sigv{6}$\\
$ \varphi_W^+ \bar \varphi_Z\ $& $W^+_\mu Z_\mu $&$\sigv{6}$ &
$ \varphi_W^-\bar \varphi_Z\ $& $ W^-_\mu Z_\mu $ & $\sigv{6}$\\

$\varphi_Z\ \bar \varphi_W^+$& $W^+_\mu A_\mu $&$\sigv{7}$ &
$\varphi_Z\ \bar \varphi_W^-$& $ W^-_\mu A_\mu $ &$\sigv{7}$ \\
$ \varphi_W^+ \bar \varphi_Z\ $& $W^+_\mu A_\mu $&$\sigv{7}$ &
$ \varphi_W^-\bar \varphi_Z\ $& $ W^-_\mu A_\mu$ & $\sigv{7}$\\

$\varphi_W^-\ \bar \varphi_W^+$& $A_\mu A_\mu$ &$\sigv{8}$ &
$\varphi_W^+\ \bar \varphi_W^-$& $A_\mu A_\mu$ & $\sigv{8}$\\
$\varphi_W^-\ \bar \varphi_W^+$& $Z_\mu A_\mu$ &$\sigv{9}$ &
$\varphi_W^+\ \bar \varphi_W^-$& $Z_\mu A_\mu$ &$\sigv{9}$ \\
$\varphi_W^-\ \bar \varphi_W^+$& $Z_\mu Z_\mu$ &$\sigv{10}$ &
$\varphi_W^+\ \bar \varphi_W^-$& $Z_\mu Z_\mu$ &$\sigv{10}$ \\
$\varphi_W^-\ \bar \varphi_W^+$& $W^+_\mu W^-_\mu$ &$\sigv{11}$ &
$\varphi_W^+\ \bar \varphi_W^-$& $W^+_\mu W^-_\mu$ &$\sigv{11}$ \\
$\varphi_W^+\ \bar \varphi_W^+$& $W^+_\mu W^+_\mu$ &$\sigv{12}$ &
$\varphi_W^-\ \bar \varphi_W^-$& $W^-_\mu W^-_\mu$ &$\sigv{12}$ \\

\end{tabular}
\caption{Coannihilation processes with vector final states}
\end{table}

\subsection{Higgs Final States}

\begin{table}[!h]
\centering
\begin{tabular}{c|c|c||c|c|c}
Initial state & Final state & $\langle \sigma v \rangle$ & Initial
state & Final state & $\langle \sigma v \rangle$\\\hline

$\varphi_W^-\ \bar \varphi_W^+$& $H H$ & $\sigv{13}$ &
$\varphi_W^+\ \bar \varphi_W^-$& $H H$ &$\sigv{13}$ \\
$\varphi_Z\ \bar \varphi_Z$& $H H$ &$\sigv{14}$ \\

\end{tabular}
\caption{Coannihilation processes for Higgs final states}
\end{table}

\subsection{Fermionic Final States}
We define $f$ as the set of fermions in the Standard Model ($u,\ c,\
t,\ d,\ s,\ b,\ e^-,\ \mu^-,\ \tau^-, \nu_e,\ \nu_\mu,\ \nu_\tau$) and
$\bar f$ the set of antifermions. 
\begin{table}[!h]
\centering
\begin{tabular}{c|c|c||c|c|c}
Initial state & Final state & $\langle \sigma v \rangle$ & Initial state & Final state & $\langle \sigma v \rangle$\\\hline
$\varphi_A\ \bar \varphi_A$& $f\bar f$ &$\sigv{15}$ &
$\varphi_A\ \bar \varphi_Z$& $f\bar f$ & $\sigv{16}$\\
$\varphi_A\ \bar \varphi_W^+$& $f\bar f$ &$\sigv{17}$ &
$\varphi_A\ \bar \varphi_W^-$& $f\bar f$ &$\sigv{17}$ \\
$\varphi_Z\ \bar \varphi_A$& $f\bar f$ &$\sigv{16}$ &
$\varphi_W^+\ \bar \varphi_A$& $f\bar f$ &$\sigv{17}$ \\
$\varphi_W^-\ \bar \varphi_A$& $f\bar f$ &$\sigv{17}$ &
$\varphi_Z\ \bar \varphi_Z$& $f\bar f$ &$\sigv{18}$ \\
$\varphi_W^+\ \bar \varphi_Z$& $f\bar f$ &$\sigv{19}$ &
$\varphi_W^-\ \bar \varphi_Z$& $f\bar f$ & $\sigv{19}$\\
$\varphi_Z\ \bar \varphi_W^+$& $f\bar f$ &$\sigv{19}$ &
$\varphi_Z\ \bar \varphi_W^-$& $f\bar f$ &$\sigv{19}$ \\
$\varphi_W^-\ \bar \varphi_W^+$& $f\bar f$ &$\sigv{20}$ &
$\varphi_W^+\ \bar \varphi_W^-$& $f\bar f$ & $\sigv{20}$\\

\end{tabular}
\caption{Coannihilation processes with fermionic final states}
\end{table}

\bibliographystyle{toolfiles/JHEP}
\bibliography{content/biblio}

\providecommand{\href}[2]{#2}\begingroup\raggedright\begin{thebibliography}{10}

\bibitem{Nordstrom:1988fi}
G.~Nordstrom, {\it {On the possibility of unifying the electromagnetic and the
  gravitational fields}},  {\em Phys. Z.} {\bf 15} (1914) 504--506,
  [\href{http://arxiv.org/abs/physics/0702221}{{\tt physics/0702221}}].

\bibitem{Kaluza:1921tu}
T.~Kaluza, {\it {On the Problem of Unity in Physics}},  {\em Sitzungsber.
  Preuss. Akad. Wiss. Berlin (Math. Phys.)} {\bf 1921} (1921) 966--972.

\bibitem{Klein:1926tv}
O.~Klein, {\it {Quantum Theory and Five-Dimensional Theory of Relativity. (In
  German and English)}},  {\em Z. Phys.} {\bf 37} (1926) 895--906. [Surveys
  High Energ. Phys.5,241(1986)].

\bibitem{Servant2003a}
G.~Servant and T.~M.~P. Tait, {\it {Is the lightest Kaluza-Klein particle a
  viable dark matter candidate?}},  {\em Nuclear Physics B} {\bf 650} (2003),
  no.~1-2 391--419, [\href{http://arxiv.org/abs/hep-ph/0206071}{{\tt
  hep-ph/0206071}}].

\bibitem{Antoniadis:1990ew}
I.~Antoniadis, {\it {A Possible new dimension at a few TeV}},  {\em Phys.
  Lett.} {\bf B246} (1990) 377--384.

\bibitem{Antoniadis:1998ig}
I.~Antoniadis, N.~Arkani-Hamed, S.~Dimopoulos, and G.~R. Dvali, {\it {New
  dimensions at a millimeter to a Fermi and superstrings at a TeV}},  {\em
  Phys. Lett.} {\bf B436} (1998) 257--263,
  [\href{http://arxiv.org/abs/hep-ph/9804398}{{\tt hep-ph/9804398}}].

\bibitem{Appelquist2000}
T.~Appelquist, H.-C. Cheng, and B.~a. Dobrescu, {\it {Bounds on universal extra
  dimensions}},  {\em Physical Review D} {\bf 64} (June, 2001) 035002,
  [\href{http://arxiv.org/abs/hep-ph/0012100}{{\tt hep-ph/0012100}}].

\bibitem{Randall:1999vf}
L.~Randall and R.~Sundrum, {\it {An Alternative to compactification}},  {\em
  Phys. Rev. Lett.} {\bf 83} (1999) 4690--4693,
  [\href{http://arxiv.org/abs/hep-th/9906064}{{\tt hep-th/9906064}}].

\bibitem{Agashe:2007jb}
K.~Agashe, A.~Falkowski, I.~Low, and G.~Servant, {\it {KK Parity in Warped
  Extra Dimension}},  {\em JHEP} {\bf 04} (2008) 027,
  [\href{http://arxiv.org/abs/0712.2455}{{\tt arXiv:0712.2455}}].

\bibitem{Cacciapaglia:2009pa}
G.~Cacciapaglia, A.~Deandrea, and J.~Llodra-Perez, {\it {A Dark Matter
  candidate from Lorentz Invariance in 6D}},  {\em Journal of High Energy
  Physics} {\bf 03} (2010) 083, [\href{http://arxiv.org/abs/0907.4993}{{\tt
  arXiv:0907.4993}}].

\bibitem{Dobrescu:2001ae}
B.~A. Dobrescu and E.~Poppitz, {\it {Number of fermion generations derived from
  anomaly cancellation}},  {\em Physical Review Letters} {\bf 87} (2001)
  031801, [\href{http://arxiv.org/abs/hep-ph/0102010}{{\tt hep-ph/0102010}}].

\bibitem{Maru2010}
N.~Maru, T.~Nomura, J.~Sato, and M.~Yamanaka, {\it {The universal extra
  dimensional model with S2/Z2 extra-space}},  {\em Nuclear Physics B} {\bf
  830} (May, 2010) 414--433, [\href{http://arxiv.org/abs/0904.1909}{{\tt
  arXiv:0904.1909}}].

\bibitem{Dohi2010}
H.~Dohi and K.~ya~Oda, {\it {Universal extra dimensions on real projective
  plane}},  {\em Physics Letters B} {\bf 692} (Apr., 2010) 114--120,
  [\href{http://arxiv.org/abs/1004.3722}{{\tt arXiv:1004.3722}}].

\bibitem{RandjbarDaemi:1982hi}
S.~Randjbar-Daemi, A.~Salam, and J.~A. Strathdee, {\it {Spontaneous
  Compactification in Six-Dimensional Einstein-Maxwell Theory}},  {\em Nuclear
  Physics B} {\bf 214} (1983) 491--512.

\bibitem{Frere:2003}
J.~M. Fr\`ere, M.~V. Libanov, E.~Y. Nugaev, and S.~V. Troitsky, {\it {Fermions
  in the vortex background on a sphere}},  {\em JHEP} {\bf 06} (2003) 009,
  [\href{http://arxiv.org/abs/hep-ph/0304117}{{\tt hep-ph/0304117}}].

\bibitem{Frere:2013}
J.~M. Fr\`ere, M.~Libanov, S.~Mollet, and S.~Troitsky, {\it {Neutrino hierarchy
  and fermion spectrum from a single family in six dimensions: realistic
  predictions}},  {\em JHEP} {\bf 08} (2013) 078,
  [\href{http://arxiv.org/abs/1305.4320}{{\tt arXiv:1305.4320}}].

\bibitem{Frere:2015}
J.-M. Fr\`ere, M.~Libanov, S.~Mollet, and S.~Troitsky, {\it {Flavour changing
  $Z'$ signals in a 6D inspired model}},
  \href{http://arxiv.org/abs/1505.08017}{{\tt arXiv:1505.08017}}.

\bibitem{Winslow:2010nk}
P.~T. Winslow, K.~Sigurdson, and J.~N. Ng, {\it {Multi-State Dark Matter from
  Spherical Extra Dimensions}},  {\em Phys. Rev.} {\bf D82} (2010) 023512,
  [\href{http://arxiv.org/abs/1005.3013}{{\tt arXiv:1005.3013}}].

\bibitem{flurry1980symmetry}
R.~Flurry, {\em Symmetry groups: theory and chemical applications}.
\newblock Prentice-Hall, 1980.

\bibitem{RauschdeTraubenberg:2005aa}
M.~Rausch~de Traubenberg, {\it {Clifford algebras in physics}},  in {\em {7th
  International Conference on Clifford Algebras and their Applications in
  Mathematical Physics Toulouse, France, May 19-29, 2005}}, 2005.
\newblock \href{http://arxiv.org/abs/hep-th/0506011}{{\tt hep-th/0506011}}.

\bibitem{Dobrescu:2004zi}
B.~A. Dobrescu and E.~Ponton, {\it {Chiral compactification on a square}},
  {\em Journal of High Energy Physics} {\bf 03} (2004) 071,
  [\href{http://arxiv.org/abs/hep-th/0401032}{{\tt hep-th/0401032}}].

\bibitem{berline1992heat}
A.~Lichnerowicz, {\it {Champs spinoriels et propagateurs en relativit\'e
  g\'en\'erale}},  {\em Bulletin de la Socit\'et\'e Math\'ematique de France}
  {\bf 92} (1964) 11.

\bibitem{Conway2002}
J.~H. Conway and D.~H. Huson, {\it {The Orbifold Notation for Two-Dimensional
  Groups}},  {\em Structural Chemistry} {\bf 13} (2002), no.~3.

\bibitem{Conway2008}
J.~H. Conway, H.~Burgiel, and C.~Goodman-Strauss, {\em {The Symmetries of
  Things}}.
\newblock A. K. Peters, Ltd, Wellesley, Massachusetts, 2008.

\bibitem{Carroll}
S.~M. Carroll and M.~M. Guica, {\it {Sidestepping the cosmological constant
  with football shaped extra dimensions}},
  \href{http://arxiv.org/abs/hep-th/0302067}{{\tt hep-th/0302067}}.

\bibitem{Geraci2008}
A.~A. Geraci, S.~J. Smullin, D.~M. Weld, J.~Chiaverini, and A.~Kapitulnik, {\it
  {Improved constraints on non-Newtonian forces at 10 microns}},  {\em Physical
  Review D} {\bf 78} (feb, 2008) 12,
  [\href{http://arxiv.org/abs/0802.2350}{{\tt arXiv:0802.2350}}].

\bibitem{Gelfand1958}
I.~M. Gelfand, R.~A. Minlos, and Z.~J. Shapiro, {\em {Representations of the
  rotation group and of the Lorentz group, and their applications}}.
\newblock MacMillan, 1963.

\bibitem{Newman1966}
E.~T. Newman and R.~Penrose, {\it {Note on the Bondi-Metzner-Sachs Group}},
  {\em Journal of Mathematical Physics} {\bf 7} (1966), no.~5 863.

\bibitem{Penrose1984}
R.~Penrose and W.~Rindler, {\em {Spinors and space-time, Vol. I}}, vol.~1.
\newblock Cambridge University Press, 1984.

\bibitem{Castillo2007}
G.~T. del Castillo, {\it {Spin-weighted spherical harmonics and their
  applications}},  {\em Revista mexicana de f\'{\i}sica} {\bf 53} (2007), no.~2
  125--134.

\bibitem{Dohi2014}
H.~Dohi, T.~Kakuda, K.~Nishiwaki, K.-y. Oda, and N.~Okuda, {\it {Notes on
  sphere-based universal extra dimensions}},
  \href{http://arxiv.org/abs/1406.1954}{{\tt arXiv:1406.1954}}.

\bibitem{Camporesi1996}
R.~Camporesi and A.~Higuchi, {\it {On the eigenfunctions of the Dirac operator
  on spheres and real hyperbolic spaces}},  {\em Journal of Geometry and
  Physics} {\bf 20} (1996).

\bibitem{Abrikosov2002}
A.~A. Abrikosov, {\it {Dirac operator on the Riemann sphere}},
  \href{http://arxiv.org/abs/hep-th/0212134}{{\tt hep-th/0212134}}.

\bibitem{Nilse2006}
L.~Nilse, {\it {Classification of 1D and 2D Orbifolds}},
  \href{http://arxiv.org/abs/hep-ph/0601015}{{\tt hep-ph/0601015}}.

\bibitem{Cheng2002}
H.~C. Cheng, K.~T. Matchev, and M.~Schmaltz, {\it {Radiative corrections to
  Kaluza-Klein masses}},  {\em Physical Review D} {\bf 66} (Apr., 2002)
  hep--ph/036005, [\href{http://arxiv.org/abs/hep-ph/0204342}{{\tt
  hep-ph/0204342}}].

\bibitem{Cacciapaglia2011}
G.~Cacciapaglia, A.~Deandrea, and J.~Llodra-Perez, {\it {The universal Real
  Projective Plane: LHC phenomenology at one loop}},  {\em Journal of High
  Energy Physics} (Apr., 2011) 45, [\href{http://arxiv.org/abs/1104.3800}{{\tt
  arXiv:1104.3800}}].

\bibitem{DaRold2004}
L.~{Da Rold}, {\it {Radiative corrections in 5D and 6D expanding in winding
  modes}},  {\em Physical Review D} {\bf 69} (2004), no.~10 1--18,
  [\href{http://arxiv.org/abs/0311063}{{\tt 0311063}}].

\bibitem{Cacciapaglia2012}
G.~Cacciapaglia and B.~Kubik, {\it {Even tiers and resonances on the real
  projective plane}},  {\em Journal of High Energy Physics} {\bf 2013} (feb,
  2013) 52, [\href{http://arxiv.org/abs/1209.6556}{{\tt arXiv:1209.6556}}].

\bibitem{ATLASCollaboration2014}
{ATLAS Collaboration}, {\it {Search for high-mass dilepton resonances in pp
  collisions at $\sqrt{s}=\tevm{8}$ with the ATLAS detector}},  {\em Physical
  Review D} {\bf 90} (May, 2014) 18,
  [\href{http://arxiv.org/abs/1405.4123}{{\tt arXiv:1405.4123}}].

\bibitem{CMSCollaboration2014}
{CMS Collaboration}, {\it {Search for physics beyond the standard model in
  dilepton mass spectra in proton-proton collisions at $ \sqrt{s}=8 $ TeV}},
  {\em Journal of High Energy Physics} {\bf 1504} (2015) 025,
  [\href{http://arxiv.org/abs/1412.6302}{{\tt arXiv:1412.6302}}].

\bibitem{Alloul2013}
A.~Alloul, N.~D. Christensen, C.~Degrande, C.~Duhr, and B.~Fuks, {\it
  {FeynRules 2.0 - A complete toolbox for tree-level phenomenology}},  {\em
  Computer Physics Communications} {\bf 185} (Oct., 2014) 2250--2300,
  [\href{http://arxiv.org/abs/1310.1921}{{\tt arXiv:1310.1921}}].

\bibitem{Alwall2014}
J.~Alwall, R.~Frederix, S.~Frixione, V.~Hirschi, F.~Maltoni, O.~Mattelaer,
  H.~S. Shao, T.~Stelzer, P.~Torrielli, and M.~Zaro, {\it {The automated
  computation of tree-level and next-to-leading order differential cross
  sections, and their matching to parton shower simulations}},
  \href{http://arxiv.org/abs/1405.0301}{{\tt arXiv:1405.0301}}.

\bibitem{Andersson2006}
B.~Andersson, {\it {PYTHIA 6.4 Physics and Manual}},
  \href{http://arxiv.org/abs/hep-ph/0603175}{{\tt hep-ph/0603175}}.

\bibitem{DeFavereau2014}
J.~{De Favereau}, C.~Delaere, P.~Demin, a.~Giammanco, V.~Lema\^{\i}tre,
  a.~Mertens, and M.~Selvaggi, {\it {DELPHES 3: A modular framework for fast
  simulation of a generic collider experiment}},  {\em Journal of High Energy
  Physics} {\bf 2014} (2014), no.~2 [\href{http://arxiv.org/abs/0903.2225}{{\tt
  arXiv:0903.2225}}].

\bibitem{Dumont2014}
B.~Dumont, B.~Fuks, S.~Kraml, S.~Bein, G.~Chalons, E.~Conte, S.~Kulkarni,
  D.~Sengupta, and C.~Wymant, {\it {Towards a public analysis database for LHC
  new physics searches using MadAnalysis 5}},
  \href{http://arxiv.org/abs/1407.3278}{{\tt arXiv:1407.3278}}.

\bibitem{Conte2012}
E.~Conte, B.~Fuks, and G.~Serret, {\it {MadAnalysis 5, a user-friendly
  framework for collider phenomenology}},  {\em Computer Physics
  Communications} {\bf 184} (Jan., 2013) 222--256,
  [\href{http://arxiv.org/abs/1206.1599}{{\tt arXiv:1206.1599}}].

\bibitem{Conte2014}
E.~Conte, B.~Dumont, B.~Fuks, and C.~Wymant, {\it {Designing and recasting LHC
  analyses with MadAnalysis 5}},  {\em European Physical Journal} {\bf C74}
  (May, 2014) 3103, [\href{http://arxiv.org/abs/1405.3982}{{\tt
  arXiv:1405.3982}}].

\bibitem{Nussinov:1985xr}
S.~Nussinov, {\it {Technocosmology: Could a Technibaryon Excess Provide a
  ``Natural'' Missing Mass Candidate?}},  {\em Phys. Lett.} {\bf B165} (1985)
  55.

\bibitem{Petraki:2013wwa}
K.~Petraki and R.~R. Volkas, {\it {Review of asymmetric dark matter}},  {\em
  Int. J. Mod. Phys.} {\bf A28} (2013) 1330028,
  [\href{http://arxiv.org/abs/1305.4939}{{\tt arXiv:1305.4939}}].

\bibitem{Zurek:2013wia}
K.~M. Zurek, {\it {Asymmetric Dark Matter: Theories, Signatures, and
  Constraints}},  {\em Phys. Rept.} {\bf 537} (2014) 91--121,
  [\href{http://arxiv.org/abs/1308.0338}{{\tt arXiv:1308.0338}}].

\bibitem{Griest1991}
K.~Griest and D.~Seckel, {\it {Three exceptions in the calculation of relic
  abundances}},  {\em Physical Review D} {\bf 43} (1991), no.~10 3191--3203.

\bibitem{Arbey2012a}
A.~Arbey, G.~Cacciapaglia, A.~Deandrea, and B.~Kubik, {\it {Dark Matter in a
  twisted bottle}},  {\em Journal of High Energy Physics} {\bf 2013} (Oct.,
  2013) 43, [\href{http://arxiv.org/abs/1210.0384}{{\tt arXiv:1210.0384}}].

\bibitem{Belyaev}
A.~Belyaev, N.~D. Christensen, and A.~Pukhov, {\it {CalcHEP 3.4 for collider
  physics within and beyond the Standard Model}},  {\em Computer Physics
  Communications} {\bf 184} (2013), no.~7 1729--1769,
  [\href{http://arxiv.org/abs/1207.6082}{{\tt arXiv:1207.6082}}].

\bibitem{WolframResearchInc}
{Wolfram Resarch Inc.}, {\it {Mathematica 8}},  2010.

\bibitem{Kong2005}
K.~Kong and K.~T. Matchev, {\it {Precise Calculation of the Relic Density of
  Kaluza-Klein Dark Matter in Universal Extra Dimensions}},  {\em Journal of
  High Energy Physics} {\bf 2006} (2006), no.~01 038,
  [\href{http://arxiv.org/abs/0509119}{{\tt 0509119}}].

\bibitem{PlanckCollaboration2015}
{Planck Collaboration}, {\it {Planck 2015 results. XIII. Cosmological
  parameters}},  \href{http://arxiv.org/abs/1502.01589}{{\tt
  arXiv:1502.01589}}.

\end{thebibliography}\endgroup


\providecommand{\href}[2]{#2}\begingroup\raggedright\endgroup


\providecommand{\href}[2]{#2}\begingroup\raggedright\endgroup

\end{document}